\documentclass{aa}  

\usepackage{amsmath}
\usepackage{graphicx}
\usepackage{txfonts}
\usepackage{natbib}
\bibpunct{(}{)}{;}{a}{}{,} % to follow the A&A style

\begin{document}

\title{SOMBI: Bayesian identification of parameter relations in unstructured cosmological data}

\author{Philipp Frank\inst{\ref{inst2},\ref{inst3}}
\and Jens Jasche \inst{\ref{inst1}}
\and Torsten A. En\ss lin \inst{\ref{inst2},\ref{inst3}}}

\institute{
Max-Planck Institut für Astrophysik, Karl-Schwarzschild-Str. 1, 85748, Garching, Germany \label{inst2}
\and Ludwig-Maximilians-Universität München, Geschwister-Scholl-Platz 1, 80539, München, Germany\label{inst3}
\and Excellence Cluster Universe, Technische Universit\"at M\"unchen, Boltzmannstrasse 2, 85748 Garching, Germany \label{inst1}}

%\date{Received 9 Mai 2014 / Accepted x Mai 2014}
%\date{Received 08 September 2014}

\abstract{This work describes the implementation and application of a correlation determination method based on Self Organizing Maps and Bayesian Inference (SOMBI). SOMBI aims to automatically identify relations between different observed parameters in unstructured cosmological or astrophysical surveys by automatically identifying data clusters in high-dimensional datasets via the Self Organizing Map neural network algorithm. Parameter relations are then revealed by means of a Bayesian inference within respective identified data clusters. Specifically such relations are assumed to be parametrized as a polynomial of unknown order. The Bayesian approach results in a posterior probability distribution function for respective polynomial coefficients. To decide which polynomial order suffices to describe correlation structures in data, we include a method for model selection, the Bayesian Information Criterion, to the analysis. The performance of the SOMBI algorithm is tested with mock data. As illustration we also provide applications of our method to cosmological data. In particular, we present results of a correlation analysis between galaxy and AGN properties provided by the SDSS catalog with the cosmic large-scale-structure (LSS). The results indicate that the combined galaxy and LSS dataset indeed is clustered into several sub-samples of data with different average properties (for example different stellar masses or web-type classifications). The majority of data clusters appear to have a similar correlation structure between galaxy properties and the LSS. In particular we revealed a positive and linear dependency between the stellar mass, the absolute magnitude and the color of a galaxy with the corresponding cosmic density field. A remaining subset of data shows inverted correlations, which might be an artifact of non-linear redshift distortions.}

\maketitle

\section{Introduction}\label{sec:intro}
Coming decades will witness an avalanche of cosmological data generated by new astronomical facilities such as the LSST \citep[see][]{2009arXiv0912.0201L}, SKA \citep[see][]{2004NewAR..48..979C} or the spaceborne Euclid mission \citep[see][]{2011arXiv1110.3193L}. These new generation of telescopes will produce enormous amounts of unstructured data. Unlike laboratory experiments on Earth, which perform targeted searches for specific physical processes, cosmological surveys always record a combination of several different, but interacting phenomena as well as systematics. Challenges for coming data analyses therefore arise from the requirement to handle such unstructured datasets in order to either test current physical theories or identify new phenomena.  

Generally cosmological or astronomical observations are generated by a combination of different physical effects. As an example galaxies and stars form in deep gravitational potential wells of the underlying dark matter distribution. The depth and shape of such potentials can affect the formation of galaxies and stars and therefore the emission spectra of photons that we observe. Consequently, correlations between the properties of galaxies and their large scale environment provide insights into the mechanisms of galaxy formation \citep[see e.g.][]{1974ApJ...194....1O,1980ApJ...236..351D,1984ApJ...281...95P,2001ApJ...557..117B,2003ApJ...584..210G,1996MNRAS.283..709H,2002MNRAS.334..673L,2005ApJ...629..143B,2004MNRAS.353..713K,2003AAS...202.5103H,LEELI2008,2011MNRAS.415.1797C,2015ApJ...812....4Y,2016MNRAS.456..571R}. When propagating through the Universe, photons will be affected by cosmic expansion, dust extinction, and absorption in the intergalactic medium, yielding altered spectra when detected at the telescope. This simple example manifests that cosmological observations generally detect a combination of several different physical phenomena which we need to separate in order to identify and study individual aspects of our physical theories.

Separating the observations into different sub groups, which permit to study such individual aspects, is a particular challenging task in regimes of huge datasets or in high dimensions where manual clustering of data is not feasible. As a response to this problem we present in this work a Bayesian inference approach to automatically search for data clusters and relations between observed parameters without human intervention. In addition the Bayesian approach permits to provide corresponding uncertainty quantification for all inferred parameters.
  
Specifically we develop a Bayesian procedure to identify deterministic relations between pairs of observable parameters. In doing so we assume the relation between parameters to be described by an arbitrarily non-linear function which can be Taylor expanded to any given polynomial order.
By restricting our analysis to a Taylor expansion of arbitrary polynomial order we show that the task of identifying non-linear relations between parameters becomes a linear inference problem. To automatically and self-consistently identify the optimal polynomial order which is supported by the available data, we use the Bayesian Information Criterion (BIC) \citep[see e.g.][and references therein]{2007MNRAS.377L..74L}. To further handle unstructured data consisting of a combination of many different processes generating the observations we use a specific type of artificial neural network to separate individual data clusters. In the past, various implementations of neural networks have been used in order to structure cosmological data \citep[see e.g.][]{1997astro.ph..4012N,2013A&A...559A...7F,2012MNRAS.419.2633G,2016MNRAS.455.4289L,1995ApJ...452L..77M,2015IAUGA..2258115P}.

In particular we will use so called self organizing maps \citep[SOM, first presented by][]{Kohonen1982}, which map the topology of a high dimensional data space to a hyper space of lower dimensionality. As a consequence SOMs are an ideal tool to separate individual data clusters and to perform studies of the relations between parameters in such clusters. 

In this work we present theory and implementation details of our method based on Self Organizing Maps and Bayesian Inference (SOMBI) as well as several detailed tests of it.

As illustrative examples we also apply the SOMBI algorithm to a galaxy and an active galactic nuclei (AGN) catalog to study relations between the properties of observed galaxies and the large scale cosmic environment. Specifically we combine galaxy properties provided by the Sloan Digital Sky Survey (SDSS) with large scale structure properties derived from reconstructed density fields provided by \citet[][]{JLW15}. This results in a combined dataset permitting to perform a detailed correlation analysis between galaxies, AGNs and the large scale environment hosting them.

The results obtained from our tests and data applications demonstrate that SOMBI is a powerful tool to study unstructured datasets in an astrophysical or cosmological setting.

In Section \ref{sec:met} we describe the methodology to identify correlations between observed parameters and demonstrate the ability of SOMs to separate clusters in unstructured datasets.
In Section \ref{sec:data} we present the datasets used for correlation analysis as well as the density field reconstructions provided by \citet[][]{JLW15}.
In order to reveal the correlation structure of galaxies and the cosmic large-scale-structure (LSS), in Section \ref{sec:daapp} we apply the SOMBI algorithm to data and discuss the results. In order to verify our methodology, we compare it to results obtained by \citet{LEELI2008} as well as state of the art methods for sub-division of cosmological data. 
Finally, in Section \ref{sec:summ}, we conclude the paper by discussing our results.

\section{Methodology}\label{sec:met}
This work presents a method to study relations between different parameters in unstructured cosmological or astrophysical observations. As a show case in Section \ref{sec:daapp} we will study correlations between SDSS galaxy properties and the LSS.

Detection as well as interpretation of physical quantities from raw data is a very complex task and generally requires broad expert knowledge. A particular challenge arises from the requirement to capture all systematics and involved assumptions in an accurate data model to infer physical quantities. The goal of this work is to present a generic method able to reveal correlations in complex datasets, which does not require human intervention.

In general, datasets consist of a combination of data drawn from multiple generation processes. This results in sub-samples of data holding various different correlation structures. In order to infer correlations form data correctly, we have to disentangle samples corresponding to independent data generation processes.

To do so, we assume that data-spaces used for correlation determination consist of sub-samples of data drawn from multiple simple data generation processes instead of one complex process. It is important to point out that we assume that each sub-sample is drawn from one generation process and that the correlations corresponding to one process can be modeled by unique correlation functions. Therefore a major goal of this work is to present methods able to reveal sub-samples regarding these assumptions. 
Within such a sub-sample we have to identify the correlation structure and model it parametrically.

\subsection{Parametric model}\label{sec:param}
The goal of correlation analysis is to find the correlation structure between two quantities $x$ and $y$. In order to model the correlation function we assume that the relation between $x$ and $y$ can be written as:
\begin{equation} \label{eq:mode}
y=f(x)+n
\end{equation}
where $f$ is an arbitrary unknown function and $n$ is assumed to be uncorrelated, normal distributed noise. The underlying assumption of this relation is that $x$ has a causal impact on $y$. If it were the other way around, $x$ and $y$ have to be interchanged.

If $f$ is assumed to be continuously differentiable then it can be expanded in a Taylor series up to $M$th order and equation \eqref{eq:mode} yields:
\begin{equation} \label{eq:mod}
y\approx \sum\limits_{i=0}^{M} f_i x^i + n \ .
\end{equation}

Determination of correlations therefore requires to determine optimal coefficients $f_i$ for a given set of $U$ data points $d_i = (x_i,y_i), i\in[1,...,U]$. Eq. \eqref{eq:mod} should hold for every data point in the sample and therefore results in $U$ relations which can be recombined into a linear system of equations by defining vectors $\mathbf{y} := (y_1 , y_2 , ... , y_U)^T$ and $\mathbf{f} := (f_0 , f_1 , ... , f_M)^T$. Specifically,
\begin{align}\label{eq:setup}
\mathbf{y} &=  \mathbf{R} \mathbf{f} +\mathbf{n} = \notag\\ 
\begin{pmatrix}
y_1\\y_2\\...\\y_U
\end{pmatrix} &= \begin{pmatrix}
x_1^0 & x_1^1&x_1^2&...&x_1^M\\
...&...&...&...&...\\
x_U^0&x_U^1&x_U^2&...&x_U^M
\end{pmatrix}  \begin{pmatrix}
f_0\\f_1\\...\\f_M
\end{pmatrix}+\begin{pmatrix}
n_1\\n_2\\...\\n_U
\end{pmatrix} \ .
\end{align}
Without further knowledge about the noise we assume $\mathbf{n}$ to obey Gaussian statistics with zero mean and diagonal covariance. This assumes the noise of individual data points to be uncorrelated. We further add the restriction that each $n_i$ has the same variance $p$. This is reasonable if there are no locally varying uncertainties in the data space. Therefore the probability distribution for $\mathbf{n}$ is defined as:
\begin{equation}\label{eq:n}
P(\mathbf{n}|\mathbf{N}) := \mathcal{G}(\mathbf{n},\mathbf{N}) = \frac{1}{|2 \pi \mathbf{N}|^{\frac{1}{2}}} e^{-\frac{1}{2} \mathbf{n}^T\mathbf{N}^{-1}\mathbf{n}}
\end{equation}
where $N_{ij}=p \ \delta_{ij}$ and $|\mathbf{N}| = p^U$ denotes the determinant of $\mathbf{N}$. Since it is required that $p \geq 0$, it can be parametrized as $p := e^{\eta}$, where the unknown constant $\eta \in \mathbb{R}$ needs to be inferred from the data.

The goal of this method is to identify parameter relations in unstructured datasets. In a generic case, the probability distribution of the noise $P(\mathbf{n})$ might differ from Eq. \eqref{eq:n}. In particular the noise can be correlated. If the correlation structure is known, this can easily be encoded in $\mathbf{N}$. In this case the formal procedure of inference presented in the following is still valid, but the solutions (also the possibility of finding exact solutions) strongly depend on the form of $P(\mathbf{n})$.
	
In contrast, in a more generic case the correlation structure of $\mathbf{n}$ can be unknown. In this case, correlations are indistinguishable of the correlation structure of the ``signal'' $\mathbf{Rf}$ due to the fact that $\mathbf{Rf}$ and $\mathbf{n}$ contribute to the data $\mathbf{y}$ in the same way (as indicated in Eq. \eqref{eq:setup}). Therefore the choice of $P(\mathbf{n}|\mathbf{N})$ as given in Eq. \eqref{eq:n} is still acceptable, but the interpretation of the revealed correlation function $f(x)$ changes. In particular $f(x)$ represents the correlation between $x$ and $y$ as well as
the correlation structure of $\mathbf{n}$. A final interpretation of such correlations may require additional information on the noise properties to disentangle noise from signal. However, this is a fundamental requirement of any method aiming and inferring signals from observations. A more general, non-Gaussian noise case would require a more substantial extension of SOMBI.

The prior distribution for $\eta$ is assumed to be flat because a priori, the noise could have any magnitude, and therefore no value for $\eta$ should be preferred. The joint probability of $\mathbf{f}$, $\mathbf{d}$ and $\eta$ can be obtained by marginalization over $\mathbf{n}$ and use of the data model given in Eq. \eqref{eq:setup}. We further assume the prior on $\mathbf{f}$ to be flat to permit $\mathbf{f}$ to model an arbitrary polynomial of order $M$. This yields:
\begin{align}\label{eq:jointp}
P(\mathbf{f},\mathbf{d},\eta)  &=  \int P(\mathbf{f},\mathbf{d},\eta,\mathbf{n}) \ \mathrm{d}\mathbf{n}   \notag\\ &= \int P(\mathbf{d}|\mathbf{f},\eta,\mathbf{n}) \ P(\mathbf{f}) \  P(\mathbf{n}|\eta) \  P(\eta) \ \mathrm{d}\mathbf{n}   \notag\\ &\propto \int \delta^D(\mathbf{y}-(\mathbf{R}\mathbf{f}+\mathbf{n})) \  \mathcal{G}(\mathbf{n},\mathbf{N}) \ \mathrm{d}\mathbf{n}  \notag\\ &= \mathcal{G}(\mathbf{y}-\mathbf{R}\mathbf{f},\mathbf{N}) =  \frac{1}{|2 \pi \mathbf{N}|^{\frac{1}{2}}} e^{-\frac{1}{2} (\mathbf{y}-\mathbf{R}\mathbf{f})^T\mathbf{N}^{-1}(\mathbf{y}-\mathbf{R}\mathbf{f})} \ .
\end{align}
Completing the square in the exponent, Eq. \eqref{eq:jointp} yields:
\begin{equation}\label{eq:jointp2}
P(\mathbf{f},\mathbf{d},\eta) \propto \frac{1}{|2 \pi \mathbf{N}|^{\frac{1}{2}}} e^{-\frac{1}{2} (\mathbf{y}^T\mathbf{N}^{-1}\mathbf{y} - \mathbf{j}^T\mathbf{D}\mathbf{j})} e^{-\frac{1}{2} (\mathbf{f}-\mathbf{D}\mathbf{j})^T\mathbf{D}^{-1}(\mathbf{f}-\mathbf{D}\mathbf{j})}
\end{equation}
with $\mathbf{D}=(\mathbf{R}^T\mathbf{N}^{-1}\mathbf{R})^{-1}$ and $\mathbf{j}=\mathbf{R}^T\mathbf{N}^{-1}\mathbf{y}$.
Note that the second exponential function is a Gaussian distribution in $\mathbf{f}$ with mean $\mathbf{D}\mathbf{j}$ and covariance $\mathbf{D}$. 

The posterior probability distribution for $\mathbf{f}$ given the data $\mathbf{d}$ and the noise parameter $\eta$ can be expressed in terms of the joint probability of all quantities using Bayes theorem. Specifically: 
\begin{equation} \label{eq:probt}
P(\mathbf{f}|\mathbf{d},\eta)=\frac{P(\mathbf{f},\mathbf{d},\eta)}{P(\mathbf{d},\eta)} \ .
\end{equation}

If $\eta$ is known then the proper probability distribution of $\mathbf{f}$ given $\mathbf{d}$ and $\eta$ is obtained from Eq. \eqref{eq:jointp2} by normalization. Specifically,
\begin{equation}\label{eq:wienf}
P(\mathbf{f}|\mathbf{d},\eta)=\mathcal{G}(\mathbf{f}-\mathbf{D}\mathbf{j},\mathbf{D}) = \frac{1}{|2 \pi \mathbf{D}|^{\frac{1}{2}}} e^{-\frac{1}{2} (\mathbf{f}-\mathbf{D}\mathbf{j})^T\mathbf{D}^{-1}(\mathbf{f}-\mathbf{D}\mathbf{j})} \ ,
\end{equation}
where we ignored factors independent on $\mathbf{f}$ since the distribution is now properly normalized. Mean and covariance of this distribution are given by
\begin{align}\label{eq:fmean}
\mathbf{f}_{\mathrm{WF}} &= \left<\mathbf{f}\right>_{(\mathbf{f}|\mathbf{d},\eta)}=\mathbf{D}\mathbf{j} = (\mathbf{R}^T\mathbf{N}^{-1}\mathbf{R})^{-1} \mathbf{R}^T \mathbf{N}^{-1} \mathbf{y} \notag\\ &= (\mathbf{R}^T\mathbf{R})^{-1}\mathbf{R}^T \mathbf{y}
\end{align}
and
\begin{align}\label{eq:fcov}
\mathbf{D}&= \left<(\mathbf{f}-\left<\mathbf{f}\right>)(\mathbf{f}-\left<\mathbf{f}\right>)^T\right>_{(\mathbf{f}|\mathbf{d},\eta)} =(\mathbf{R}^T\mathbf{N}^{-1}\mathbf{R})^{-1} \notag\\ &= e^{\eta} (\mathbf{R}^T\mathbf{R})^{-1} .
\end{align}
These equations resemble the solution of a Wiener filtering equation. Note that the mean of $\mathbf{f}$ does not depend on $\eta$ since the noise is assumed to be zero centered and the prior for $\mathbf{f}$ is flat. This method determines the full posterior probability distribution for the coefficients $\mathbf{f}$.

For visualization, the posterior distribution $P(\mathbf{f}| \mathbf{d},\eta)$ (Eq. \eqref{eq:wienf}) can be transformed into data space resulting in a PDF $P(f(x)|\mathbf{d},\eta)$ for the realizations of correlation functions $f(x), \ x\in \mathbb{R}$. Specifically,
\begin{equation}\label{eq:real}
P(f(x)|\mathbf{d},\eta)=\mathcal{G}(f(x)-\mathbf{\tilde{R}}(x)\mathbf{f}_{WF},\mathbf{Y}) 
\end{equation}
with 
\begin{equation}\label{eq:conres}
\mathbf{\tilde{R}}(x) = \begin{pmatrix}
1,&x,&x^2,&...&x^M
\end{pmatrix}
\end{equation}
being the response for a continuous field $x$ and
\begin{equation}\label{eq:concov}
\mathbf{Y}_{xy} = \mathbf{\tilde{R}}(x)\mathbf{D}\mathbf{\tilde{R}}(y)^T \ .
\end{equation}
$P(f(x)|\mathbf{d},\eta)$ describes how likely a realization $f(x)$ is, given the data and $\eta$. This permits to visualize the reconstructed correlation function including corresponding uncertainties in specific areas of the data space. Details about the derivation of $P(f(x)|\mathbf{d},\eta)$ are described in Appendix \ref{sec:con}.

In order to find the true value of $\eta$, we follow the spirit of the empirical Bayes approach. In particular, we obtain $\eta$ via maximum a posteriori (MAP) estimation, given the marginal probability distribution $P(\eta|\mathbf{d})$. We assume the MAP solution $\eta_{\mathrm{MAP}}$ to be the true value for $\eta$, irregardless of possible uncertainties for the estimate of $\eta$. Given $\eta_{\mathrm{MAP}}$, we can ultimately determine the posterior distribution $P(\mathbf{f}|\mathbf{d},\eta_{\mathrm{MAP}})$ (Eq \eqref{eq:wienf}).

The marginal distribution $P(\eta|\mathbf{d})$ is obtained from Eq. \eqref{eq:jointp2} by marginalization with respect to $\mathbf{f}$:
\begin{align}
P(\eta|\mathbf{d}) &\propto P(\mathbf{d},\eta) = \int P(\mathbf{f},\mathbf{d},\eta) \ \mathrm{d}\mathbf{f}  \notag\\ &\propto  \frac{1}{|2 \pi \mathbf{N}|^{\frac{1}{2}}} \ e^{-\frac{1}{2} (\mathbf{y}^T\mathbf{N}^{-1}\mathbf{y} - \mathbf{j}^T\mathbf{D}\mathbf{j})} \int e^{-\frac{1}{2} (\mathbf{f}-\mathbf{D}\mathbf{j})^T\mathbf{D}^{-1}(\mathbf{f}-\mathbf{D}\mathbf{j})} \ \mathrm{d}\mathbf{f} \notag\\ &=
\frac{|2 \pi \mathbf{D}|^{\frac{1}{2}}}{|2 \pi \mathbf{N}|^{\frac{1}{2}}} \ e^{-\frac{1}{2} (\mathbf{y}^T\mathbf{N}^{-1}\mathbf{y} - \mathbf{j}^T\mathbf{D}\mathbf{j})}
\end{align}
and the negative logarithm of this distribution is:
\begin{align}\label{eq:hamilton}
&\mathcal{H}(\eta|\mathbf{d}) = -\ln(P(\eta|\mathbf{d})) \notag\\
&= \frac{1}{2} \left[\ln(|2 \pi \mathbf{N}|)-\ln(|2\pi \mathbf{D}|) + \mathbf{y}^T\mathbf{N}^{-1}\mathbf{y} - \mathbf{j}^T\mathbf{D}\mathbf{j}\right] + \tilde{H}_0  \notag\\ &= \frac{1}{2} \left\lbrace\left[U-(M+1)\right] \eta + e^{-\eta} \left[\mathbf{y}^T\mathbf{y} - \mathbf{y}^T\mathbf{R}(\mathbf{R}^T\mathbf{R})^{-1}\mathbf{R}^T\mathbf{y}\right]\right\rbrace +H_0 \ ,
\end{align}
where in the following we call $\mathcal{H}(\eta|\mathbf{d})$ the information Hamiltonian. Here we used the definitions of $\mathbf{D}$ and $\mathbf{j}$ and the fact that $\mathbf{N}$ is diagonal. Note that $M+1$ is the dimensionality of the signal space, the space of polynomials up to order $M$ describing the y-x correlation function $f(x)$, and $U$ the dimensionality of the data space. $H_0$ and $\tilde{H}_0$ are terms independent of $\eta$. The MAP solution for $\eta$ is given by setting the first derivative of $\mathcal{H}(\eta|\mathbf{d})$ with respect to $\eta$ to zero:
\begin{align}
0 &\overset{!}{=} \dfrac{\partial \mathcal{H}(\eta|\mathbf{d})}{\partial \eta}  \notag\\ &= \frac{1}{2} \left\lbrace \left[U-(M+1)\right] - e^{-\eta} \left[\mathbf{y}^T\mathbf{y} - \mathbf{y}^T\mathbf{R}(\mathbf{R}^T\mathbf{R})^{-1}\mathbf{R}^T\mathbf{y}\right]\right\rbrace   
\end{align}
and therefore
\begin{equation}\label{eq:MAP}
p_{*}=e^{\eta_{\mathrm{MAP}}}= \frac{\mathbf{y}^T\mathbf{y} - \mathbf{y}^T\mathbf{R}(\mathbf{R}^T\mathbf{R})^{-1}\mathbf{R}^T\mathbf{y}}{U-(M+1)} \ .
\end{equation}

The Bayesian implementation of this method is able to model the posterior PDF for correlation structures in noisy data, given a specific data model (in particular the polynomial order of the Taylor series). For optimal reconstructions, the order $M$ of the polynomial describing the signal correlation needs to be known. However, for real data application the underlying model is often not known. Especially in fields where the physical processes causing correlation are not yet understood completely, it is important to have a method which does not need to know the data model in the beginning. Therefore a possible way to infer generation processes from data are described in the next section.

\subsubsection{Bayesian Information Criterion}\label{sec:BIC}
The polynomial order M up to which correlations are modeled, should be determined by the data themselves. This decision can be regarded as a model selection, with all polynomials up to order M constitute a model and the polynomial coefficients $f_i$ and the noise parameter $\eta$ are the corresponding model parameters. In order to decide which polynomial order M is sufficient to describe the correlation structure of the data, we apply the Bayesian Information Criterion \citep[see e.g.][]{2007MNRAS.377L..74L}.

The BIC approach compares the maximum of the likelihood $P(\mathbf{d}|\mathbf{f}_{\mathrm{WF}}, \eta_{\mathrm{MAP}})$ of each model modified by the number of degrees of freedom $m$. Specifically
\begin{align}
\label{eq:BICv}
\mathrm{BIC} := - 2 \ln(P(\mathbf{d}|\mathbf{f}_{\mathrm{WF}} , \eta_{\mathrm{MAP}}))+m  \ \ln(\mathrm{dim}(\mathbf{d}))=
\notag\\=\frac{1}{p_*} (\mathbf{y}-\mathbf{R}\mathbf{f}_{\mathrm{WF}})^T(\mathbf{y}-\mathbf{R}\mathbf{f}_{\mathrm{WF}})+U\ln(p_*)+(M+2)\ln(U) \ .
\end{align}
Note that if the order of the polynomial is $M$ then $m=M+2$ since there are $M+1$ polynomial coefficients $\mathbf{f}_i$ plus the noise parameter $\eta$.

\begin{figure}[htp]
	\includegraphics[scale=0.14, angle=0]{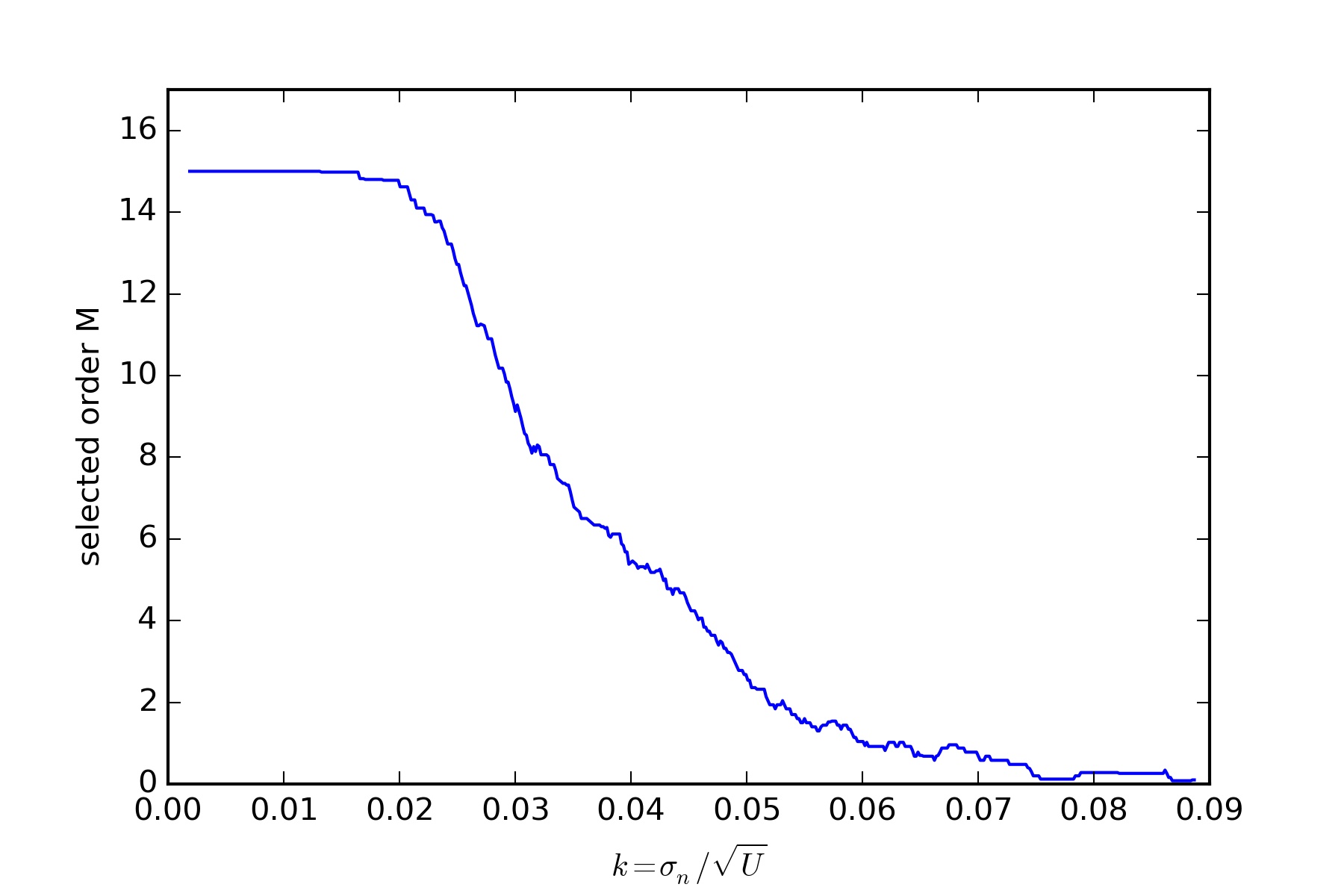}
	\centering
	\caption{Histogram of recovered polynomial order for different inverse signal to noise ratios $k$. $U=1 000$ and denotes the sample size. The noise variance $\sigma_n$ ranges from $\approx 0$ to $\leq 3$. The signal was generated according to Eq. \ref{eq:mod} as a 15th order polynomial. We see that the most adequate order selected by the BIC decreases with increasing $k$. Note that the selected model depends on the specific data realization, therefore we averaged over reconstructed orders with similar $k$}
	\label{fig:th}
\end{figure}

The application of the BIC aims to find the optimal polynomial order M that explains the observations. If the data does not support higher orders due to high impact of noise, the method will always prefer lower order polynomials even though the actual correlation might be of higher order. To demonstrate this effect we show in Fig \ref{fig:th} how the selected polynomial order M decreases with increasing noise. In the depicted case we generated mock data according to Eq. \ref{eq:mod} as a 15th order polynomial and construct samples by adding Gaussian distributed noise with different variance $\sigma_n$. In order to illustrate the impact of noise on the BIC, we depict the selected order as a function of the inverse signal to noise ratios $k=\sigma_n / \sqrt{U}$ where $U=1000$ denotes the sample size.

Combining the BIC with the parametric estimation for correlations results in a reliable method to reconstruct correlations in data and quantify corresponding uncertainties. So far we assumed data to be generated from a single process. In the following, we describe our approach to handle complex data generated by an arbitrary combination of processes.

\subsection{Self organizing maps}\label{sec:som}
The previous section describes a parametric approach to reconstruct the correlation function for noisy data drawn from a single generation process. However, real data sometimes appears to be the result from multiple generation processes with different underlying data models. This yields data samples consisting of multiple sub-samples with varying correlation structures. A successful application of the method described in Section \ref{sec:param} can only be guaranteed, if data presented to the algorithm is drawn from a single generation process. Therefore we present a method to reveal sub-samples corresponding to a single data generation process from the full dataset. In particular, we will use a Self Organizing Map.

A SOM \footnote{The SOM implementation in this work is provided by the python package PYMVPA (www.pymvpa.org/generated/ mvpa2.mappers.som.SimpleSOMMapper.html). PYMVPA is a frequently used package in computational science \citep[see e.g.][]{PYMVPA2009}} is an artificial neural network specifically designed to identify clusters in high dimensional data and to divide data into corresponding sub-samples. To accomplish this division, the SOM has to adopt the properties of the spatial distribution of data. To do so, we assume that the distribution of data points in a high-dimensional data space can be approximated by a low-dimensional manifold, mapped onto the data space. In this work, the manifold is approximated by a finite square-lattice pattern called neuron-space, consistent of a discrete set of points, called neurons. Each neuron holds a position in the neuron-space. This position is later used in order to specify links between neighboring neurons to preserve the topological structure of the approximated manifold (in this case a square lattice).

In addition to the position in the neuron-space, each neuron holds a position in data space, called weight $\mathbf{W}$. Therefore, instead of mapping a manifold onto the data space, we update the weights according to data in a way such that the chosen neuron pattern represents the data distribution. The non-linear mapping is inferred from data via a specific learning process as described in the following (for a detailed description of the algorithm see Appendix \ref{sec:apsom}).

The learning process consists of a recursive update rule, where weights get updated according to data. Data is successively presented to the network and for each iteration the ``Best Matching Unit" (BMU), which is the neuron holding the closest weight to the presented data vector $\mathbf{V}_t$, is calculated. The distance D between data vectors and weights is measured via an Euclidean metric in the normalized data space. Specifically,
\begin{equation}
	D=\sqrt{\sum\limits_{i=1}^N \left(\frac{V_i-W_i}{\sigma_i}\right)^2} \ ,
\end{equation}
where $\sigma_i$ being the scale factor for each component $i$ defined as:
\begin{equation}
	\sigma_i:=V_{i \ \mathrm{max}} - V_{i \ \mathrm{min}} \ ,
\end{equation}
where $V_{i \ \mathrm{max}}$ and $V_{i \ \mathrm{min}}$ are the maximum and minimum values of the $i$th component of all data vectors.

In addition, weights of all neurons get updated according to an update function dependent on $\mathbf{V}_t$ and the BMU. The update function,
\begin{equation}\label{eq:update}
\mathbf{W}_{t+1}=\mathbf{W}_t+ L_0 e^{-\frac{t}{\lambda}} \   \exp \left(-\frac{d_{\mathrm{BMU}}^2}{2 \sigma_t^2}\right) (\mathbf{V}_t-\mathbf{W}_t)
\end{equation}
describes the change of weights $\mathbf{W}$ after presenting the $t$th data vector $\mathbf{V}_t$ to the network. $d_{\mathrm{BMU}}$ is the distance in the neuron-space between the position of the updated neuron and the neuron identified as BMU at iteration step $t$. $L_0$ and $\lambda$ are constant tunable parameters.
$\sigma_t=\sigma_0 e^{-\frac{t}{\lambda}}$ defines the size of the neighbourhood of the BMU in the neuron-space.

Since the change of weights $\Delta W=W_{t+1} - W_t$ decreases with increasing $t$, the order of data vectors presented to the network influences the final result of the learning process. In order to avoid a bias towards data presented to the network in the beginning, we repeat the learning process multiple times for random permutations of the full dataset and average the results \citep[see][]{2001som..book.....K}.

\begin{figure*}[htp]
	
	\includegraphics[scale=0.15, angle=0]{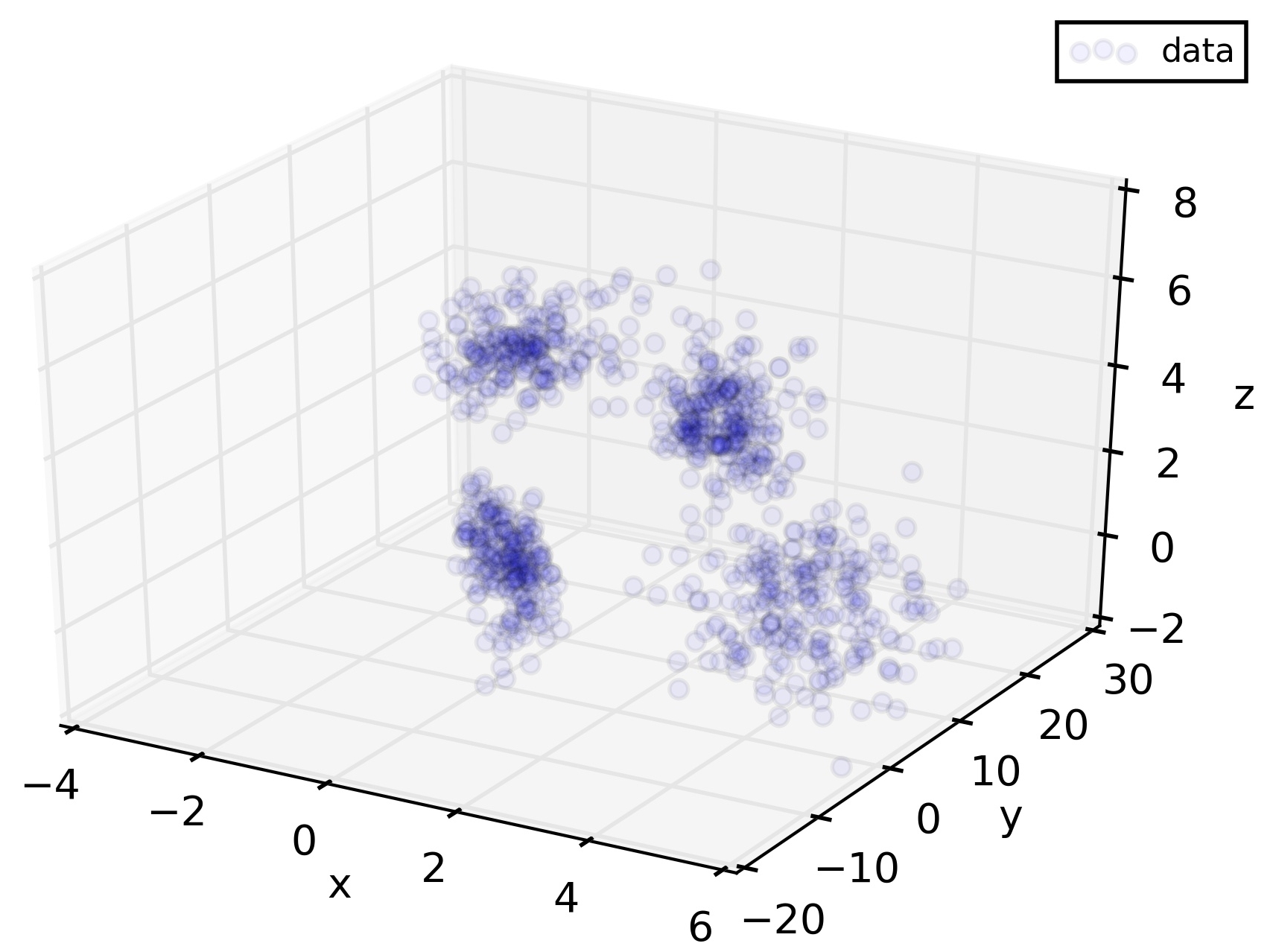}
	\includegraphics[scale=0.15, angle=0]{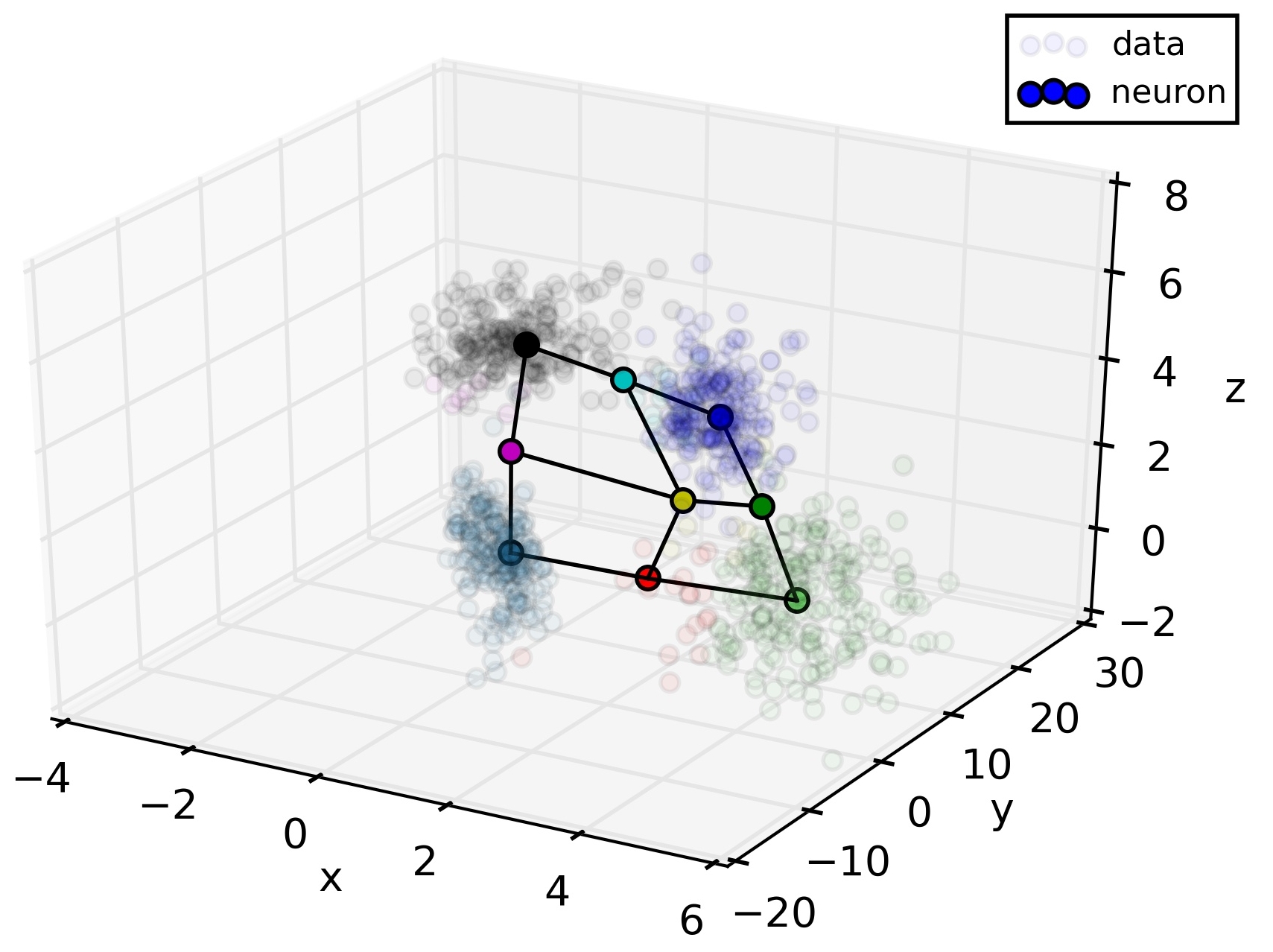}
	\centering
	\caption{The left picture shows the distribution of the mock data, generated as described in Section \ref{sec:mock}. The $x$ and $z$ coordinates for each data point are drawn from four different Gaussian distributions with different means and covariances. The covariance is assumed to be diagonal. The $y$ coordinates are generated to be correlated with the $x$ coordinates with a correlation function consistent with Eq. \eqref{eq:mod} (see Table \ref{t:mockdep} for the correlation coefficients). The right picture shows the data space including the $3 \times 3$ square lattice neuron pattern after a successful training of the SOM. Neighboring neurons are interlinked in the picture. In addition, each sub-sample of data corresponding to one neuron, is drawn in the color of the specific neuron.}
	\label{fig:damock}
\end{figure*}

The full training process can be expressed in terms of a recursive algorithm described as:

\begin{itemize}
	\item Repeat multiple times:
	\begin{itemize}
		\item Initialization of the network pattern as a square lattice.
		\item Initialization of the weights in data space for all neurons randomly.
		\item Repeat for all data vectors $\mathbf{V}_t$, $t \in (1,..,N)$:
		\begin{itemize}
			\item Calculate the BMU for $\mathbf{V}_t$, which is the closest neuron to $\mathbf{V}_t$. The distance is measured via an Euclidean metric in the normalized data space.
			\item Update the weights $\mathbf{W}$ of all neurons according to $\mathbf{V}_t$ and the BMU as described by the update function Eq. \eqref{eq:update}.
		\end{itemize}
	\end{itemize}
	\item Average the weights of each learning process for corresponding neurons.
\end{itemize}

For large datasets this training process is numerically expensive. But once completed, the trained SOM is a numerically fast and powerful tool to approximately represent the structure of datasets. A new vector $\mathbf{V}^{'}$ presented to the {\it trained} SOM is classified by the properties of the corresponding BMU. More precisely the neuron which holds the weight closest to $\mathbf{V}^{'}$ (in terms of the Euclidean distance) represents the region of the data space $\mathbf{V}^{'}$ lies in.

Regarding those properties, a SOM can be used to find data-clusters in high dimensional data spaces. Specifically, after the SOM has been trained, each training vector again is presented to the {\it trained} SOM and all vectors sharing the same BMU are stored in a sub-sample of data. Each sub-sample holds a set of data vectors with properties similar to the weight of the BMU. The average properties of this region are represented by the data space position of the BMU.

Combining the SOM approach with the parametric correlation determination results in a generic method able to identify sub-samples of data drawn from one data generation process in highly structured datasets, which we call SOMBI. In addition, the corresponding correlations for each sub-sample including a correct treatment of uncertainties is provided by SOMBI. In order to illustrate the performance of our method we apply it to mock data consistent with our data model in the following.

\subsection{Method validation with mock data}\label{sec:mock}
In this Section we demonstrate the performance of the SOMBI algorithm.

Without loss of generality in the following, we restrict the test-case to a low- (3-) dimensional mock dataset. Data vectors $\mathbf{V}$ are described by their data space positions $(x,y,z)$ in the following. The data was generated according to the data model described in Section \ref{sec:met}. Specifically, we generate a data sample consistent of 4 spatially separated sub-samples (see Figure \ref{fig:damock} left panel). Each sub-sample is Gaussian distributed among two dimensions (x- and z-axis) of the data space with varying and independent means, covariances and sample sizes for each sub-sample. In the third dimension (y-axis) of the data space we include a noisy functional dependence on x consistent with Eq. \eqref{eq:mod} for each sub-sample. The dependencies differ for each sub-sample (see Table \ref{t:mockdep} for exact correlation coefficients).

\begin{table}[h]
	\caption {Correlation coefficients between the x- and y-axis for the sub-samples of the mock data consistent with Eq. \eqref{eq:mod}}   \label{t:mockdep}
	\centering
	\begin{tabular}{ l | c c c c r }
		Sample&$f_0$&$f_1$&$f_2$ & $f_3$ & $\sigma_{\mathrm{n}}$\\\hline
		1&$1.0$ & $-4.0$ & $0.5$& $-1.0$ & $2.0$\\
		2&$-1.0$ & $0.0$ & $2.0$& & $2.0$\\
		3&$0.0$ & $1.0$ & & & $2.0$\\
		4&$3.0$ & $-2.0$ & & & $2.0$\\
	\end{tabular}
\end{table}

For optimal reconstruction each sub-sample should correspond to a single neuron of the SOM after the training process. Since the number of sub-samples for real data is not known in the beginning, we choose the number of neurons to be $9$ ($3 \times 3$ square-lattice pattern).

During training, the SOM adopts the spatial properties of the data distribution. Once completed, all data points closest to a neuron are grouped and stored in reconstructed sub-samples of data as shown in Figure \ref{fig:damock}.
As seen in the Figure, each sub-sample is now represented by one neuron located in the center of the sample. The remaining neurons get mapped between sub-samples due to the fact that the SOM aims to preserve the chosen topology for the network pattern. This results in a number of neurons holding only a tiny fraction of data without recoverable correlations. Those neurons should be excluded from further analysis. Specifically, all neurons with
\begin{equation} \label{eq:somth}
N_{\mathrm{S}} \ll \frac{N_{\mathrm{D}}}{N_{\mathrm{n}}}
\end{equation}
should be excluded. $N_{\mathrm{S}}$ denotes the number of data points corresponding to the specific neuron sample, $N_{\mathrm{D}}$ denotes the total number of data points and $N_{\mathrm{n}}$ denotes the number of neurons. The remaining neurons result in sub-samples of data (denoted as neuron-samples in the following) which represent the reconstructed data distribution.

Due to the fact that parts of the data are removed from revealed sub-samples, less signal is provided for correlation reconstruction. In particular, as the spatial separation of sub-samples corresponding to different correlation structures decreases, the amount of usable data decreases. Therefore spatial separation of sub-samples drawn from different data generation processes plays an important role for the quality of reconstructed correlation functions.

\begin{figure}[htp]
	\includegraphics[scale=0.12, angle=0]{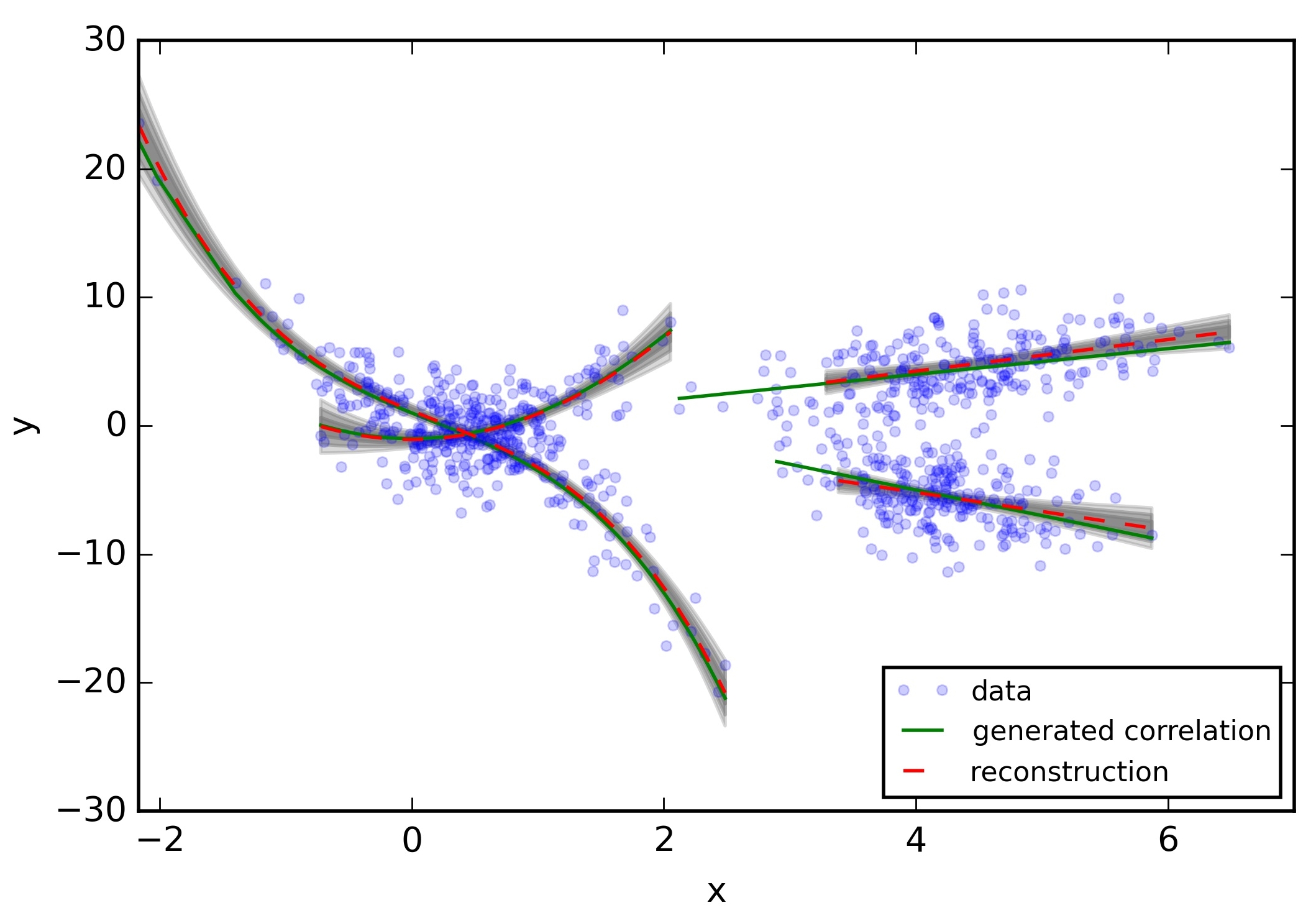}
	\centering
	\caption{The picture shows the mock data distribution projected to the $x$-$y$-plane as well as the initial correlation functions of each sub-sample of data. In addition, the reconstructed polynomials for each neuron-sample as selected after training are depicted as red dashed lines in the picture. The gray areas denote the 1-, 2- and 3-$\sigma_{\mathrm{y}}$ uncertainties of the reconstruction, respectively. $\sigma_{\mathrm{y}}$ is the projection of the parameter covariance $D$ to the data space as described by Eq. \eqref{eq:concov}.}
	\label{fig:resmock}
\end{figure}

In order to reveal the correlation structure of each neuron-sample we apply our correlation determination method to each neuron-sample. As indicated in Figure \ref{fig:resmock}, the application results in four different reconstructed correlation functions between $x$ and $y$. Each reconstructed polynomial appears to represent the initial correlation functions correctly within uncertainties. As indicated in the picture, each neuron-sample holds data corresponding to a single data generation process, allowing a successful application of the correlation determination method.

The test indicates that the method behaves as expected for consistent mock data. The SOM reveals spatially separated data clusters and the correlation determination method is valid within uncertainties. However, in this case we restrict data to consist of multiple, spatially separated, Gaussian distributed sub-samples. This restriction does not have to hold for real data. Therefore, further testing and comparison to a frequently used sub-sampling method is described in Section \ref{sec:div}. In addition, various performance tests of the SOM have been presented in literature \citep[see e.g.][]{2001som..book.....K,2012amld.book.....W}.

\subsubsection{Inconsistent mock data}
In addition to the previous example, we apply the SOMBI algorithm to a mock dataset which is inconsistent with our assumed data model. In particular we generate a two dimensional dataset $(x,y)$ with a non-polynomial correlation structure between $x$ and $y$ where we obtain $y$ by drawing a sample of a one dimensional Gaussian random field with a covariance matrix given as a diagonal matrix in Fourier space:
\begin{equation}
	P(k) = \frac{42}{(k+1)^3} \ ,
\end{equation}
where $P(k)$ is referred to as the power-spectrum.
For the purpose of this test, the exact form of the power spectrum is not crucial and was chosen for visual clarity. In order to gain a finite dataset, we discretized the function into $512$ pairs of data points $(x,y)$ (with a flat distribution in $x$). In addition we add Gaussian distributed noise to the $y$ values of the sample consistent with Eq. \eqref{eq:n}.

\begin{figure}[htp]
\includegraphics[scale=0.12, angle=0]{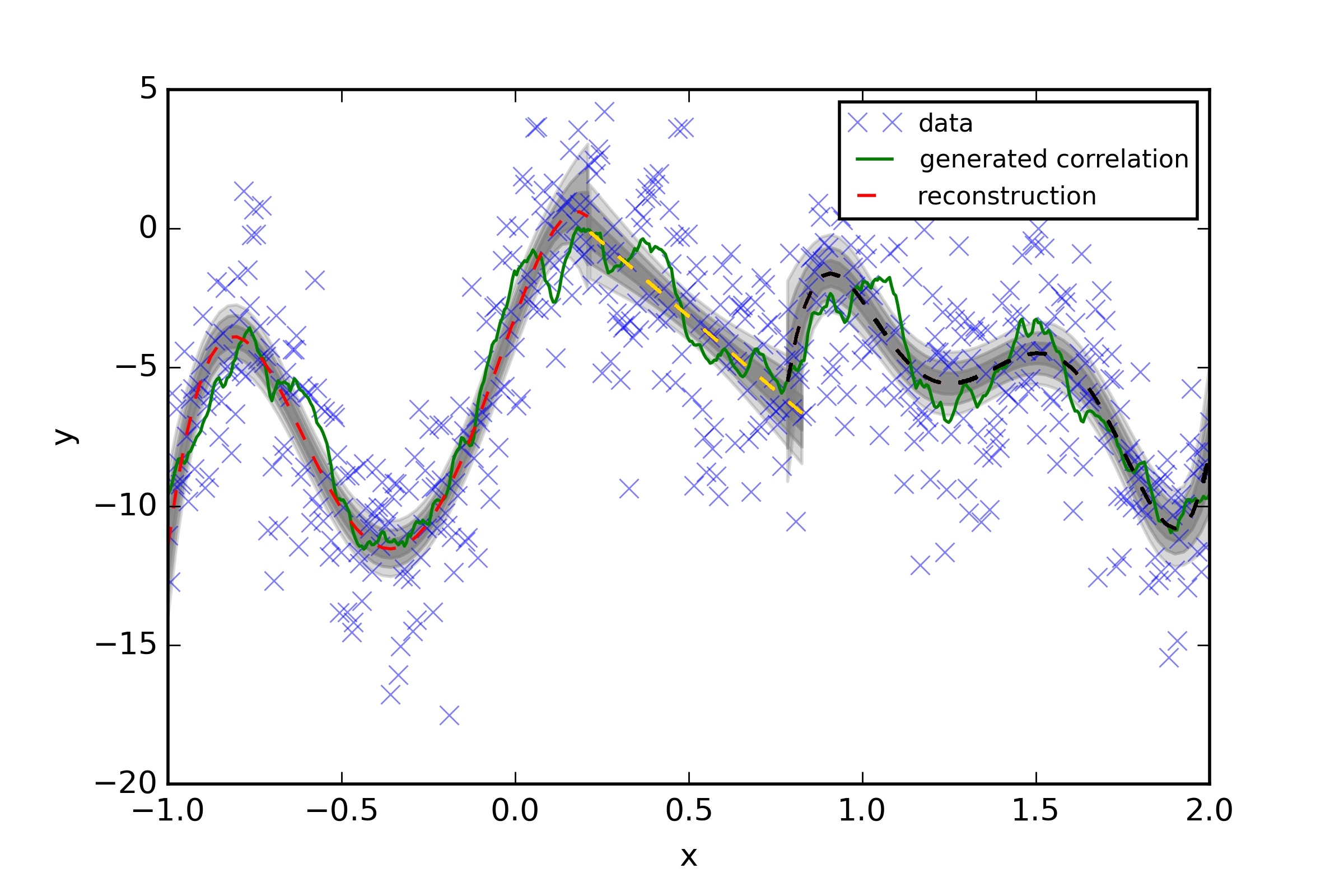}
\centering
\caption{The picture shows a correlation structure that is not of polynomial form, as indicated by the green line together with mock data generated from it. In addition the red, yellow and black dashed lines indicate the three reconstructed polynomials respectively. Gray areas denote the uncertainties of the reconstructions as described in the caption of Fig. \ref{fig:resmock}.}
\label{fig:te2}
\end{figure}

Since this dataset does not have a clustered structure, using the SOM in this context does not seem to be necessary. However, we assume that the structure of the dataset is a priori unknown. Therefore we apply the full SOMBI algorithm including the SOM to the dataset where we generate the SOM as a linear chain consistent of three neurons since the data space is only two dimensional. The application follows the same procedure as described above and results in three data constrained posterior distributions for the reconstructed polynomials. The reconstructed correlations together with the dataset and the original correlation function are depicted in Fig. \ref{fig:te2}

We see that the clustering of the SOM disentangles three sub-samples of data with structurally different reconstructed polynomials. The reconstructions support the structure of the correlation functions within uncertainties on scales where the signal dominates the noise. However, small structures in the correlations cannot be reconstructed by the algorithm since the noise dominates in those regions. In addition, the approximation of the correlation function by finite order polynomials will always result in a mismatch for non-analytic structures. However, the results support our claim that the SOM helps to disentangle complex structures into multiple but simpler structures.

\section{Data}\label{sec:data}
As a demonstration of our method, in the following we show examples of applications to galaxy data. Data used for correlation determination is described in detail in the following.

\begin{figure*}[htp]
	\includegraphics[scale=0.4, angle=0]{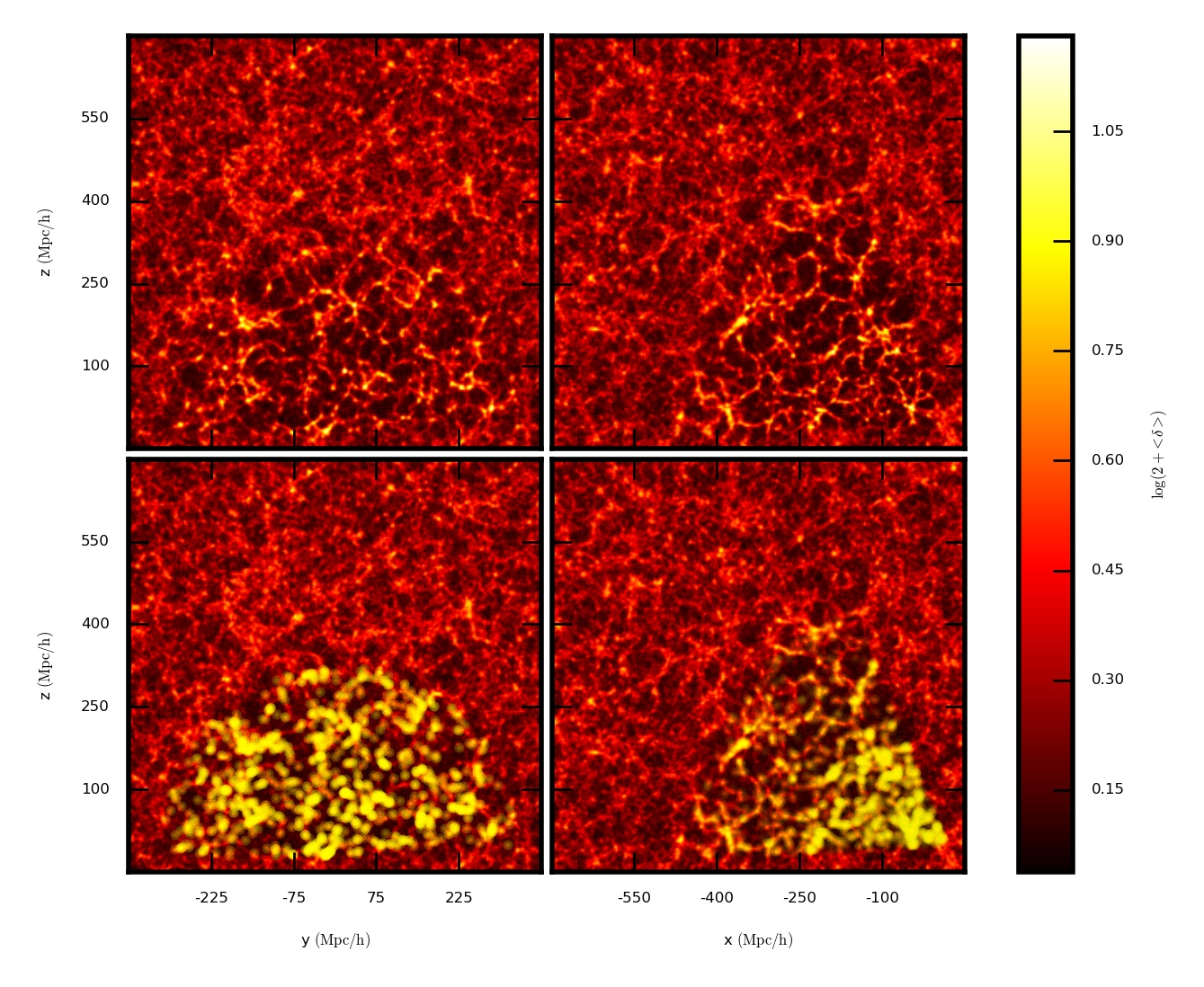}
	\centering
	\caption{Slices of the ensemble mean density field on a logarithmic scale $\log(2+\left\langle \delta \right\rangle )$ (upper panels) and the same slices with the SDSS galaxies mapped onto the grid as yellow dots (lower panels). In order to exclude areas of high uncertainty from the analysis we took a distance threshold in the co-moving frame at $d_{\mathrm{lim}}\approx 450 \ \mathrm{Mpc \ h^{-1}}$. Therefore galaxies above this limit are excluded form the analysis and not depicted.}
	\label{fig:fin}
\end{figure*}
\subsection{Galaxy data}\label{sec:sdssda}
The dataset used in this work is constructed from the sample DR7.2 of the New York University Value Added Catalog (NYU-VAGC) provided by \citet{2005AJ....129.2562B}. This catalog is based on DR7 \citep[see][]{SDSS7}, the seventh data release of the SDSS \citep[see][]{2000AJ....120.1579Y}. The sample consists of
$527 \ 372$ galaxies in total, in a redshift range of $0.001<z<0.4$. Table \ref{t:galc} shows the ranges of the catalog in the r-band Petrosian apparent magnitude $r$, the logarithm of the stellar mass $\log(M_{*})$ in units of the solar mass $M_{*}=M/M_{S} \ h^2$ and the absolute magnitude $M_{0.1 r}$. $M_{0.1 r}$ is corrected to its $z=0.1$ value according to the K-correction code of \citet{2007AJ....133..734B} and the luminosity evolution model described by \citet{2003ApJ...592..819B}. For simplicity, we restrict the application of SOMBI to these properties to look for correlations with the LSS. LSS information is provided by a set of data constrained density field reconstruction maps. Details about the reconstruction maps are described in Section \ref{sec:borg}.

\begin{table}[h]
	\caption {Property ranges of galaxy data}   \label{t:galc}
	\centering
	\begin{tabular}{ l | c c c r }
		
		&$z$&$r$&$M_{0.1r}$ & $\log(M_{*})$ \\\hline
		min&$0.001$ & $10.1$ & $-18.8$& $6.6$\\
		max&$0.4$ & $18.8$ & $-23.0$& $11.6$\\
	\end{tabular}
\end{table}

\subsection{AGN data}\label{sec:agnda}
In addition to the galaxy dataset described above, we present correlations between the LSS and an active galactic nuclei (AGN) dataset. The catalog is based on a previous SDSS data release (DR4 see \citet{2006ApJS..162...38A}) and consists of $88 \ 178$ galaxies classified as AGNs according to \cite{2003MNRAS.346.1055K}.

The data includes various properties such as luminosities of specific emission lines ([O III] 5007, [NII]) as well as stellar masses, intrinsic velocity dispersions of galaxies and parameters associated with the recent star formation history such as stellar surface mass densities and concentration indexes. For structural information, the $4000 \ \AA $ break strength is included in the dataset. In Section \ref{sec:daapp} we present the revealed correlations between each included galaxy property and the surrounding LSS. The correlation analysis is based on the method described in Section \ref{sec:met}.

\subsection{Large-scale-structure reconstructions}\label{sec:borg}
In addition to the galaxy and AGN properties directly obtained from the SDSS sample, we include properties of the cosmic LSS to our analysis and apply the SOMBI algorithm to the resulting dataset in the next section. A modern approach to LSS reconstruction is based on the reconstruction of initial conditions under the constraint of a cosmological model \citep[see e.g.][]{JLW15,JASCHEBORG2012}. The main idea of this approach lies on the fact that the initial density field follows almost homogeneous and isotropic statistics which makes a successful modeling much more likely compared to the non-linear present density field. In addition, the initial density field consists of small, very nearly Gaussian and nearly scale-invariant correlated density perturbations. Within the standard cosmology the gravitational evolution and growth of those initial perturbations, which processed the initial conditions into the present density field, is well understood in principle. As a consequence, the successful reconstruction of the initial density field ultimately results in a detailed description of the present LSS.
	
In this work we rely on results previously obtained by the BORG algorithm \citep[see][]{JASCHEBORG2012}. BORG performs reconstructions of the initial density field based on a second-order Lagrangian perturbation theory \citep[see e.g.][]{2002PhR...367....1B}. The resulting reconstructions are based on a non-linear, non-Gaussian full Bayesian LSS analysis of the SDSS DR7 main galaxy sample, the same dataset as used for our correlation study. The method used by \citet{JASCHEBORG2012} is based on a Markov Chain Monte Carlo sampling algorithm called BORG algorithm and results in a set of data constrained density contrast field samples $\delta_i \ i \in [1,..,S]$. The density contrast $\delta$ is the normalized difference of the density $\rho$ to its cosmic mean $\bar{\rho}$. Specifically,
\begin{equation}
\rho=\bar{\rho} (1+\delta) \ .
\end{equation}
The density contrast samples can be recombined to an approximate estimate for the PDF of the density contrast. Specifically,
\begin{equation}\label{eq:pdfdel}
P(\delta|\mathbf{d_*})\approx \frac{1}{S} \sum\limits_{i=1}^S \delta^D(\delta-\delta_i) \ .
\end{equation}
Applying the SOMBI methods to the density contrast samples results in a PDF describing correlation for each sample: $P(\mathbf{f}|\delta_i,\mathbf{d})$. The dependency on $\delta_i$ has to be marginalized in order to yield the final PDF $P(\mathbf{f}|\mathbf{d})$. Marginalization over $\delta$ is described in detail in Appendix \ref{sec:gm}.

\begin{figure*}[htp]
	\includegraphics[scale=0.4, angle=0]{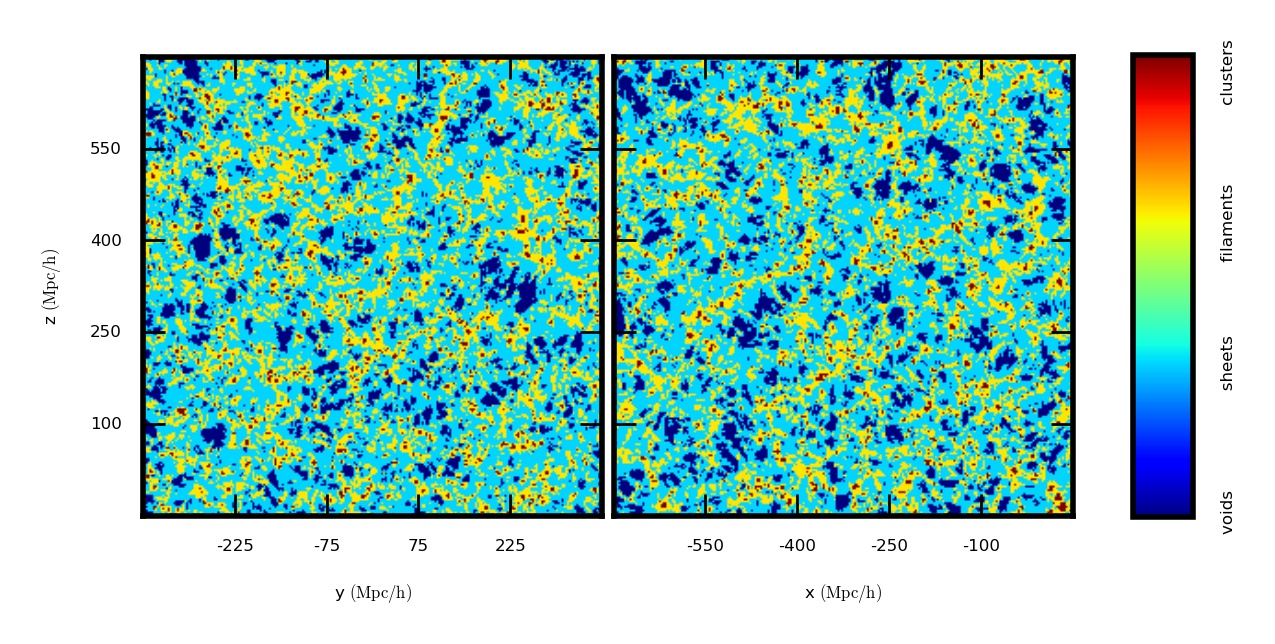}
	\centering
	\caption{Web type classification in slices of the the 3D LSS reconstruction. We cut the volume at the same positions as used in Figure \ref{fig:fin}. Since an average web type is not well defined, we present only one sample of the reconstructions instead of the mean LSS. We distinguish the LSS according to the web-type classification described in Table \ref{t:webt}. Note that sheets seem to fill the largest volume. In the chosen classification scheme, incident regions to sheets are also accounted as sheets.}
	\label{fig:fi3}
\end{figure*}

The density field inference was applied to the northern galactic cap as covered by the SDSS survey. More precisely, inference is performed on a cube with $750 \ \mathrm{Mpc \ h^{-1}}$ side length with a grid resolution of $\approx 3 \ \mathrm{Mpc \ h^{-1}}$ in the co-moving frame. This results in a cubic grid with $265^3$ voxels. Table \ref{t:bbox} denotes the boundaries of this box.

\begin{table}[h]
	\caption {Boundaries of the cubic grid in the co-moving frame}   \label{t:bbox}
	\centering
	\begin{tabular}{ c | c c }
		
		Axis & \multicolumn{2}{c}{ Boundaries ($\mathrm{Mpc \ h^{-1}}$)} \\\hline
		x & $-700$ & $50$ \\
		y&$-375$ & $375$\\
		z&$-50$ & $700$\\
	\end{tabular}
\end{table}

In order to compare galaxy properties to the properties of the LSS, we map galaxies onto the cubic grid (as depicted in Figure \ref{fig:fin}) and extract the information about the LSS provided by BORG for each position. The explicit mapping procedure is described in \ref{sec:mapping}.

The density field allows a derivation of many important quantities of the LSS. Some important examples are: the gravitational potential, the tidal-shear tensor and the web type classification.

The rescaled gravitational potential $\Phi$ is given as
\begin{equation}
\Delta \ \Phi=\delta \ 
\end{equation}
and the tidal-shear tensor $\mathbf{T}_{ij}$ is given by the Hessian of $\Phi$:
\begin{equation}
\mathbf{T}_{ij} = \dfrac{\partial^2 \Phi}{\partial x_i \partial x_j} \ .
\end{equation}
The eigenvalues of the tidal-shear tensor $\lambda_i$ (with $i \in \{1,2,3\}$), permit to classify different structure types within the cosmic matter distribution \citep[see e.g.][]{1999MNRAS.302..111L,2005MNRAS.360..216C,2006MNRAS.366.1201N}. For an application of web type classification to density fields inferred with the BORG algorithm see \citet{2015arXiv151202242L,2015JCAP...06..015L,2015A&A...576L..17L}.

In this work, we rely on the eigenvalues of the tidal-shear tensor in order to include non-local information of the LSS to the analysis. These eigenvalues provide also a coarse web type classification of the LSS in terms of voids, sheets, filaments and clusters.

\subsubsection{Web type classification of the LSS}\label{sec:web}
The web type is a classification of different structure types of the LSS. Various classification methods have been presented in literature \citep[see e.g.][]{2007A&A...474..315A,2007MNRAS.375..489H,2009MNRAS.396.1815F,2012MNRAS.425.2049H,2010MNRAS.403.1392L,2012PhRvD..85h3005S,2013MNRAS.429.1286C}. However, in this work we split the LSS into four different types (voids, sheets, filaments and clusters) according to the eigenvalues of the tidal-shear tensor following the classification procedure described by \citet{2007MNRAS.375..489H}. Table \ref{t:webt} shows the explicit classification rules and Fig. \ref{fig:fi3} shows the classification of a reconstructed sample according to these rules.

\begin{table}[htp]
	\caption {Web type classification according to the ordered eigenvalues $\lambda_1>\lambda_2>\lambda_3$ of the tidal-shear tensor. In this work we used $\lambda_{\mathrm{th}}=0$}   \label{t:webt}
	\vspace{0.2cm}
	\centering
	\begin{tabular}{ l  l  }
		\hline
		Classification &  \\\hline
		Void & $\lambda_{\mathrm{th}}>\lambda_{1},\lambda_{2},\lambda_{3} $ \\
		Sheet&$\lambda_{1}>\lambda_{\mathrm{th}}>\lambda_{2},\lambda_{3}$\\
		Filament & $\lambda_{1},\lambda_{2}>\lambda_{\mathrm{th}}>\lambda_{3}$\\
		Cluster &$\lambda_{1},\lambda_{2},\lambda_{3}>\lambda_{\mathrm{th}}$\\
		\hline
	\end{tabular}
\end{table}

The structural classification as well as the density field reconstruction itself contain information about the LSS at the location of a galaxy. These quantities are used in the following to compare galaxy properties with the LSS.

\begin{figure*}[htp]
	\includegraphics[scale=0.09, angle=0]{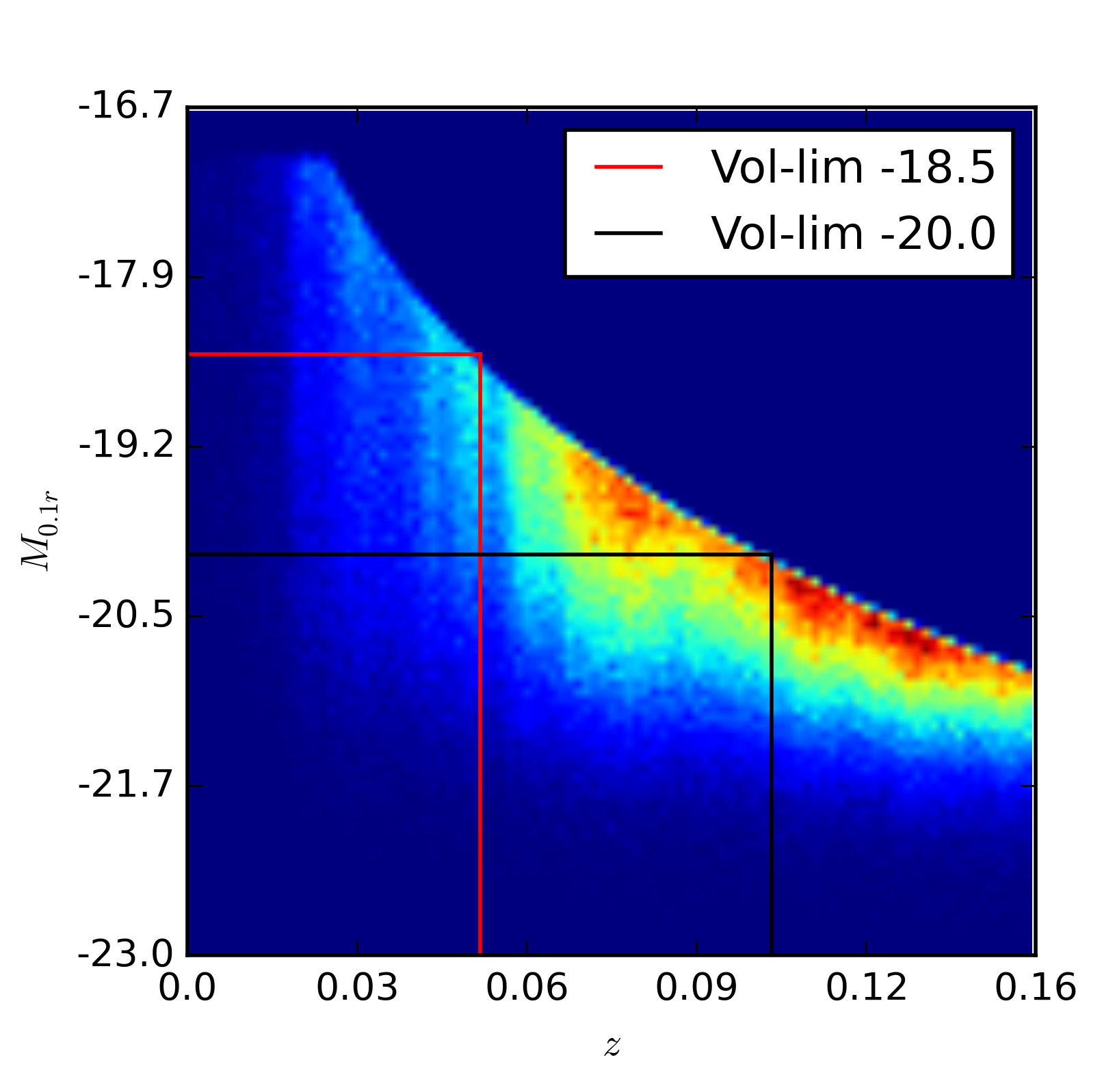}
	\includegraphics[scale=0.09, angle=0]{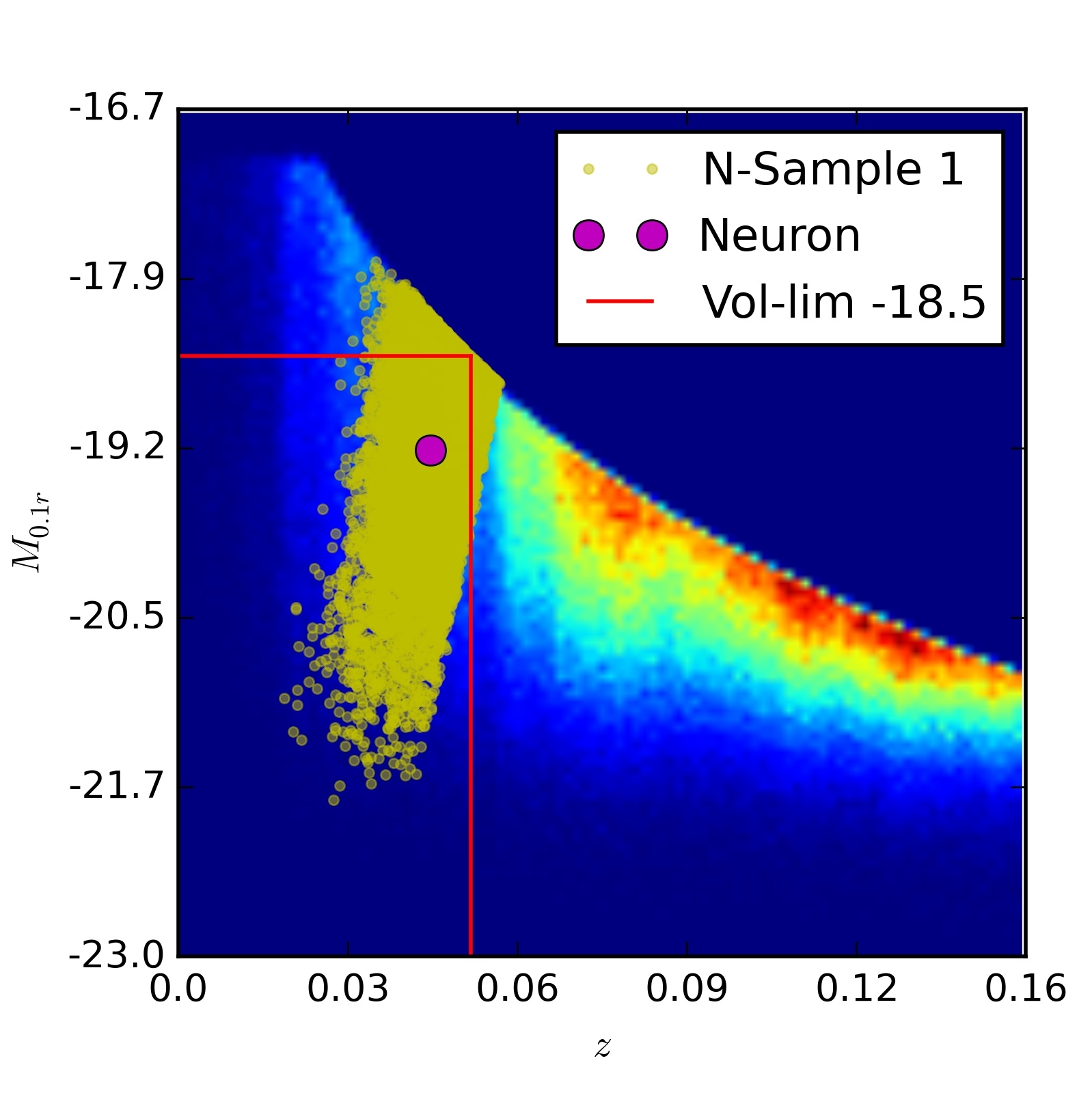}
	\includegraphics[scale=0.09, angle=0]{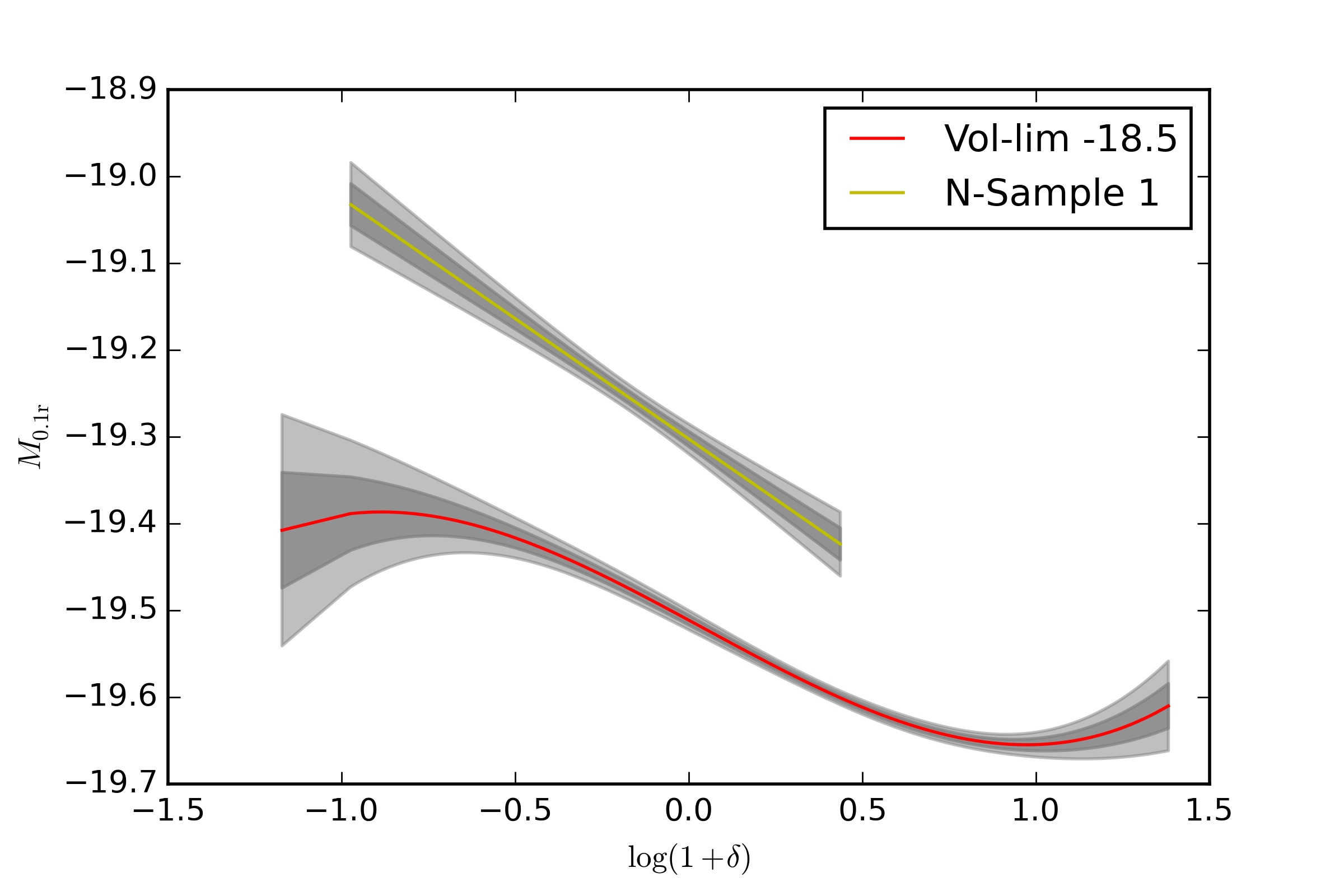}
	
	\includegraphics[scale=0.09, angle=0]{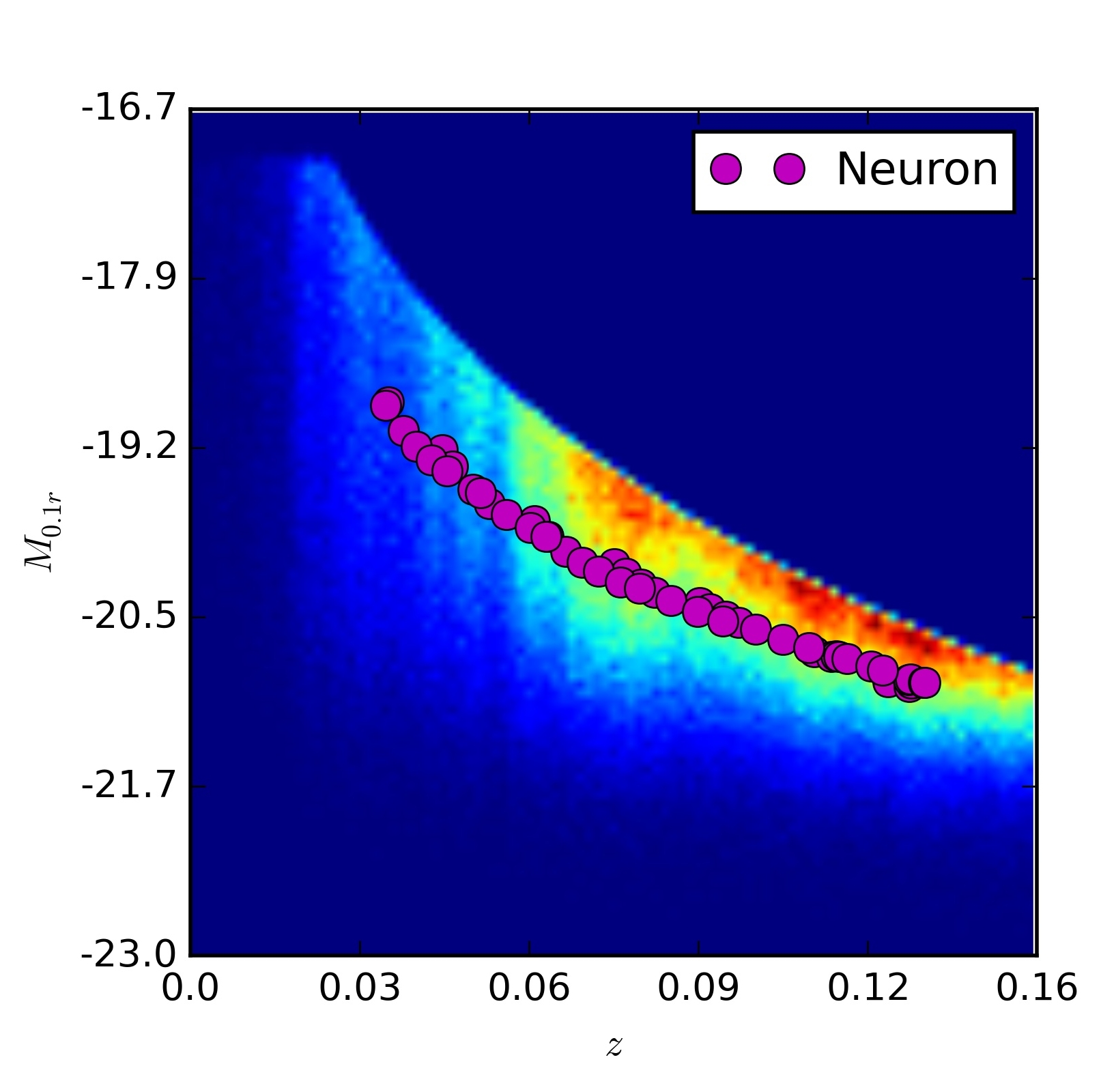}
	\includegraphics[scale=0.09, angle=0]{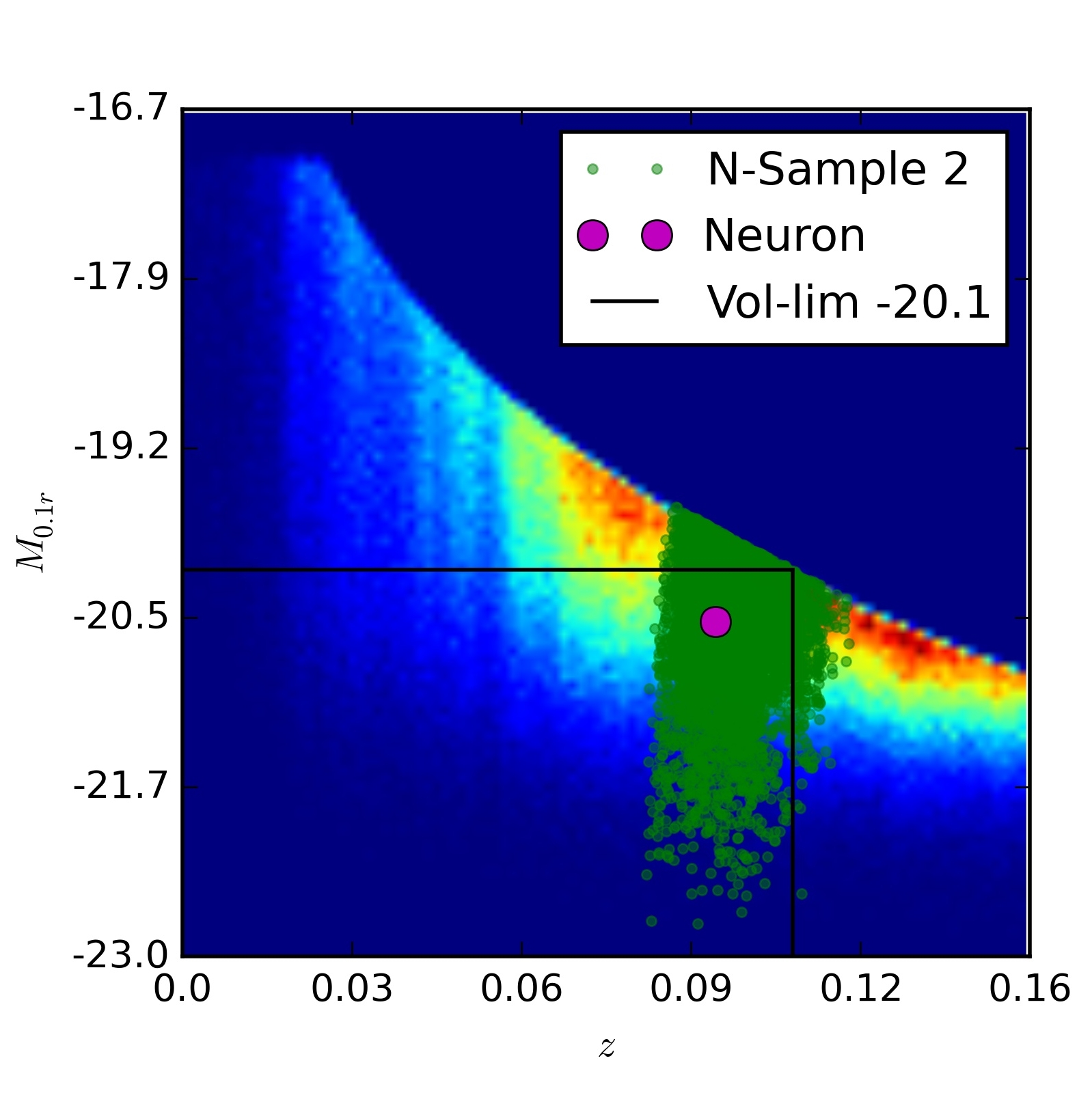}	
	\includegraphics[scale=0.09, angle=0]{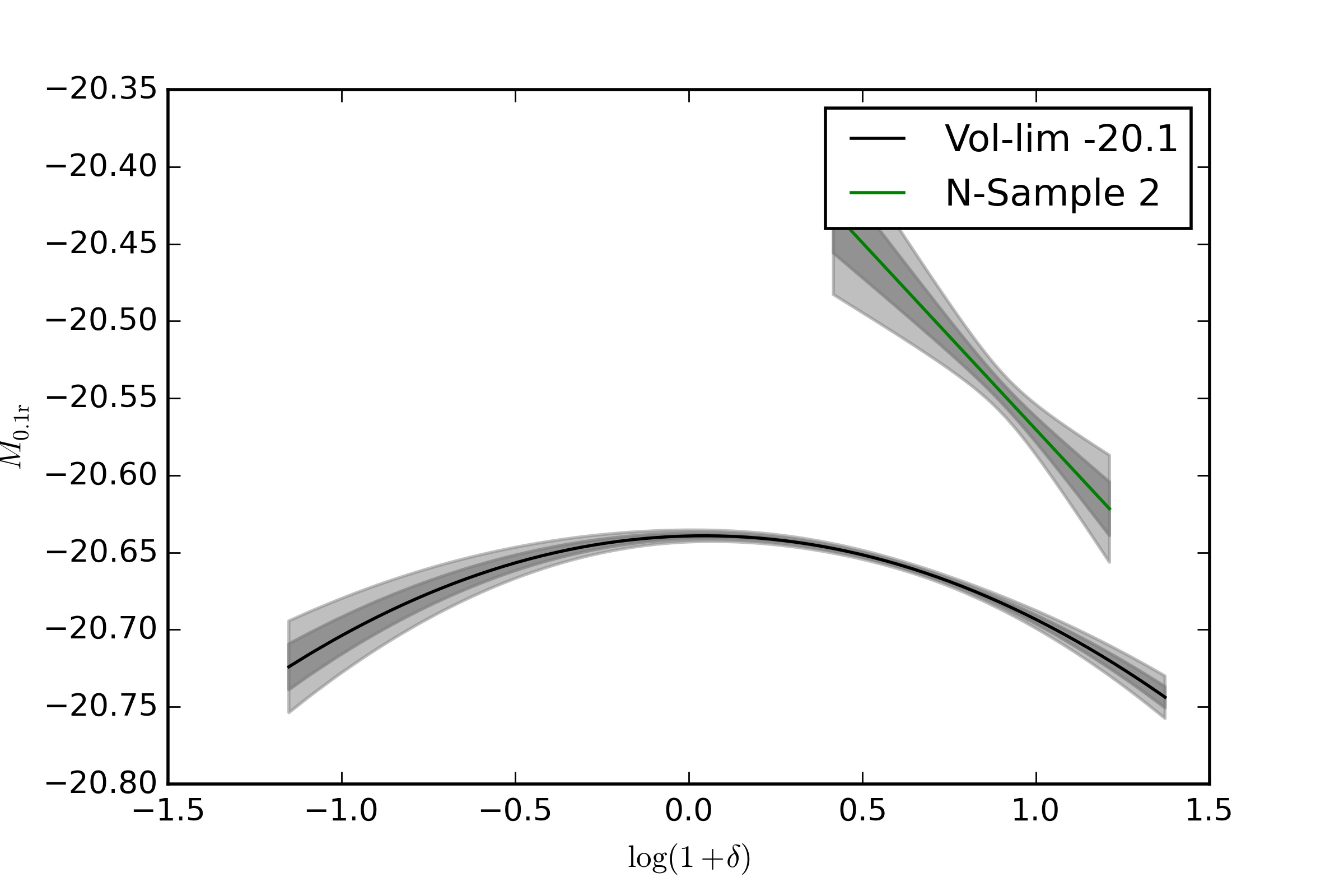}
	
	\centering
	\caption{The distribution of the SDSS data projected to the $M_{0.1 r}$-$z$-Plane. The top-left panel shows the volume limitation for different absolute magnitude thresholds ($-18.5$, $-20.0$). The bottom-left panel shows the distribution of the neurons after a successful application of the SOM to the SDSS data. The top-mid and bottom-mid panels show two different neuron-samples (N-Samples) corresponding to the depicted neurons from the trained SOM. In addition, the volume limitation used for the correlation comparison are depicted in both panels. The selected neurons hold sub-samples of data with a similar projected distribution compared to the volume-limited samples in order to compare the selection methods. The top- and bottom-right panel show reconstructed correlation functions for the volume limited sample with magnitude limits at $-18.5$ (top) and $-20.1$ (bottom) and for the corresponding neuron-samples. The range of each sub-sample in $\log(1+\delta)$ is indicated by the length of each reconstructed polynomial.}\label{fig:vol}
\end{figure*}

\section{Data application and discussion}\label{sec:daapp}
In this Section, we apply SOMBI to the galaxy and the AGN datasets derived from the SDSS survey as described in the previous Section.

In order to apply the SOM to the extended data sample we include various galaxy and LSS properties to define the data space for training. In order to find as many separated regions in data as possible, we include properties holding unique information about the data. Therefore for the SDSS catalog a reasonable setup is to include redshifts $z$, r-band absolute magnitudes $M_{0.1r}$ and colors of galaxies. To include properties of the LSS we extended the training space with the logarithm of the density field $\log(1+\delta)$ and the three eigenvalues of the tidal shear tensor at the location of each galaxy. This setup seems to be reasonable, since many properties of the LSS (for example the web type classification) depend on these quantities. The logarithm of the stellar mass $\log(M_{*})$, another common property of galaxies, was excluded from the training process since it is expected to be proportional to the absolute magnitude \citep[see e.g.][]{2005essp.book.....S,2006asco.book.....H,1997MNRAS.287..402K}. However, for correlation analysis, we will use the stellar mass again. The usage of the logarithm of the density field instead of the density field arises from the fact that the included galaxy properties are on a logarithmic scale and therefore dependencies should be estimated on this scale as well, in order to ensure an adequate data space distance measure.

\subsection{Sub-dividing the galaxy sample}\label{sec:div}
In this work we rely on the SOM to sub-divide data. However, various manual selection methods have been presented in literature \citep[see e.g.][]{2010gfe..book.....M}. In order to illustrate the performance of our method, we compare a frequently used selection method, the volume limitation, to the SOM application.

Volume limitation is an approach to account for flux limitations of telescopes. Flux limitation means that at larger distances only the brightest galaxies are detected which introduces a distance dependent selection effect onto a sample of observed galaxies. A frequently used approach to remove this effect is to limit the volume of the catalog in redshift space such that in this sub-sample all existing galaxies are included. A possible way to accomplish volume limitation is to include only galaxies to the sample brighter than a certain absolute magnitude limit $M_{\mathrm{lim}}$ and below a certain redshift limit $z_{\mathrm{lim}}$. Here $z_{\mathrm{lim}}$ is the distance at which a galaxy with absolute magnitude $M_{\mathrm{lim}}$ has an apparent magnitude equal to the survey limit $m_{\mathrm{\mathrm{lim}}}$. More precisely:
\begin{equation}\label{eq:mo}
M_{\mathrm{lim}} = m_{\mathrm{lim}}-5\log\left( \frac{r_{\mathrm{lim}}}{r_0}\right) 
\end{equation}
with $r_{\mathrm{lim}}$ being the luminosity distance corresponding to $z_{\mathrm{lim}}$ and $r_0=10 \ \mathrm{pc}$ conventionally \citep[see e.g.][]{2010gfe..book.....M}.

Figure \ref{fig:vol} shows different volume limitations of the SDSS data sample and the corresponding reconstructed correlation functions between the absolute magnitude $M_{0.1 r}$ and the logarithm of the density field $\log(1+\delta)$.

\begin{figure*}
	\includegraphics[scale=0.6, angle=0]{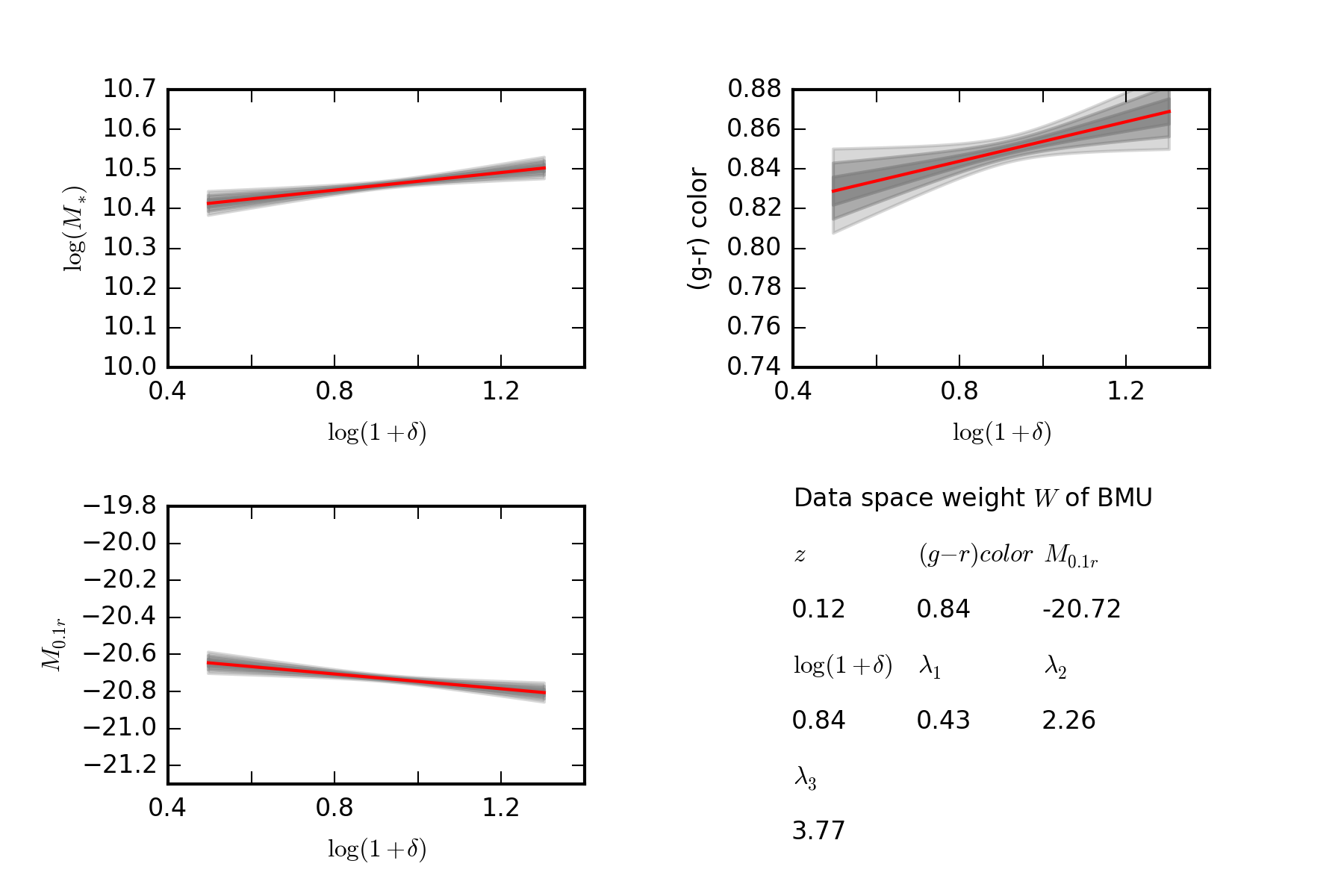}
	\includegraphics[scale=0.6, angle=0]{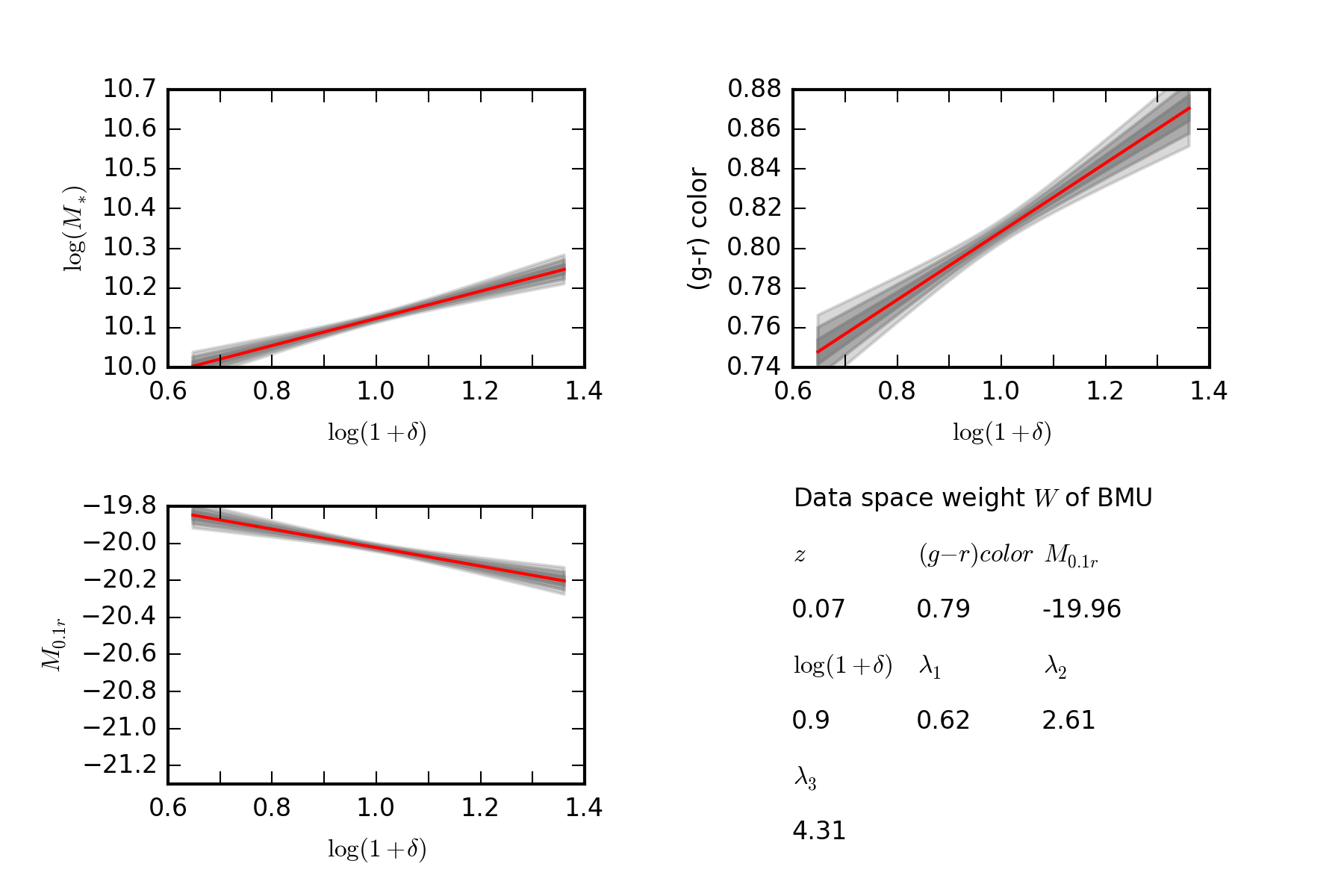}
	\includegraphics[scale=0.6, angle=0]{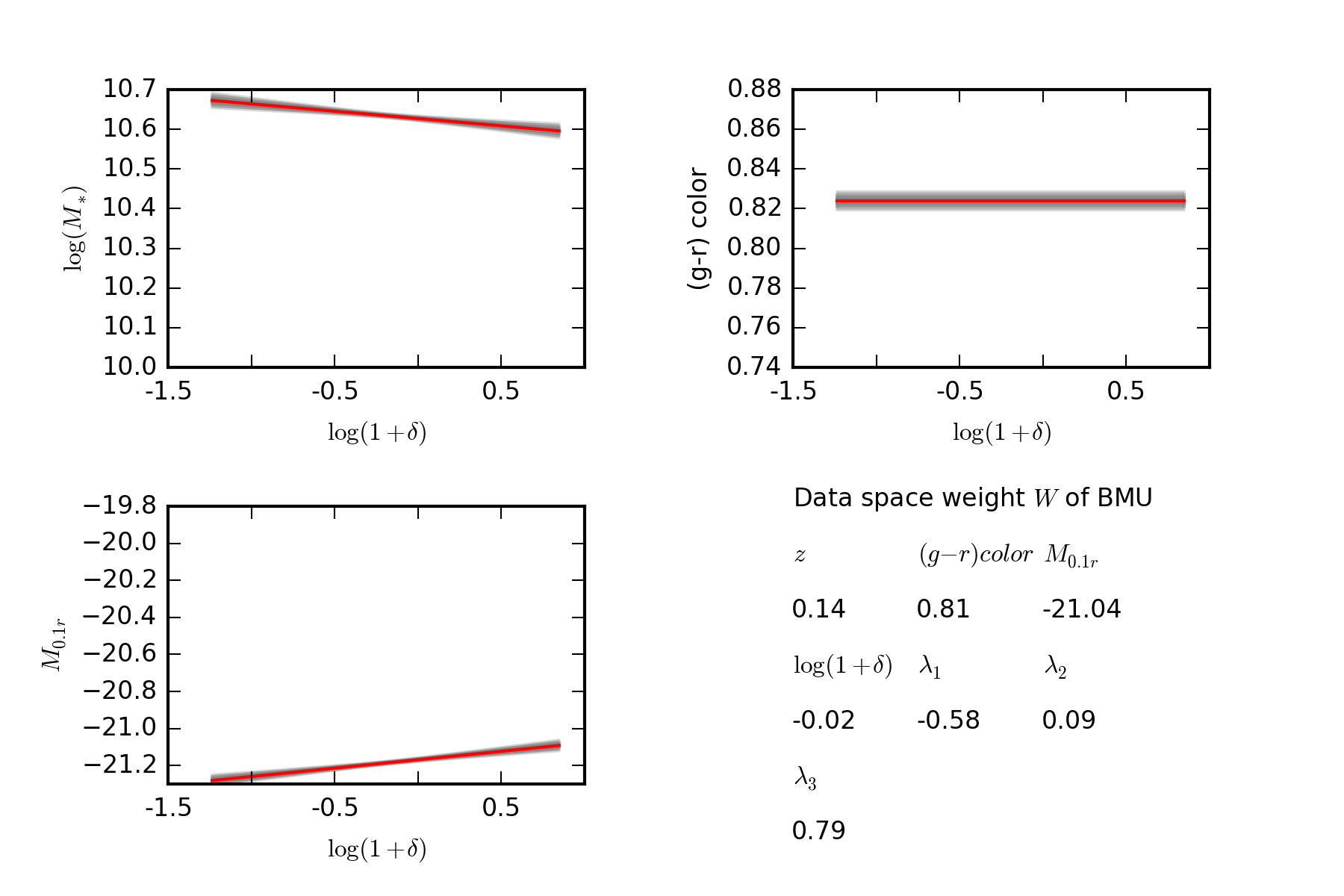}
	\includegraphics[scale=0.6, angle=0]{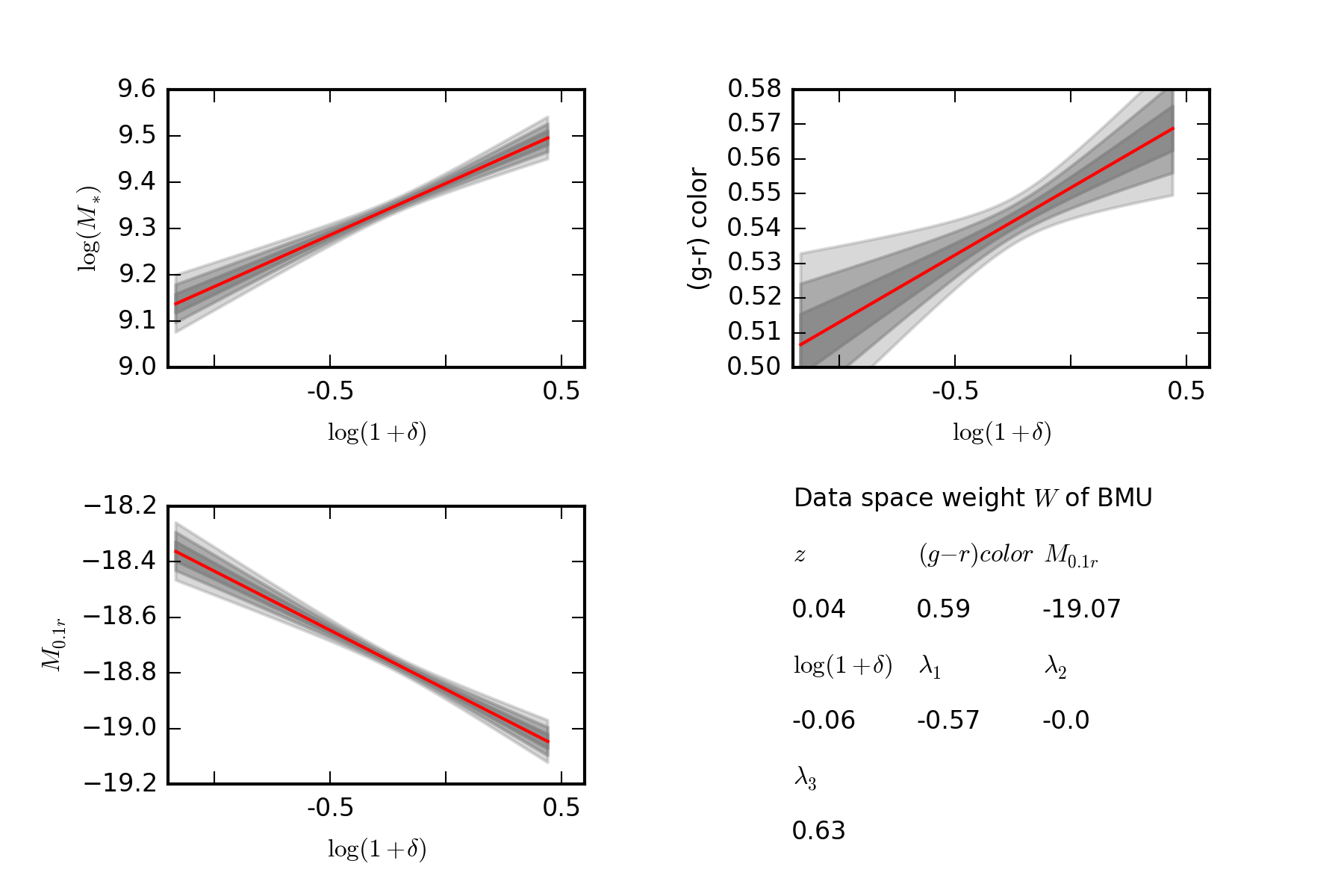}
	\centering
	\caption{Reconstructed correlation functions for different neuron-samples selected from the SDSS data sample by the SOM. In particular we depict the correlations for the logarithm of the stellar mass $\log(M_{*})$, the r-band absolute magnitude $M_{0.1 r}$ and the g-r color. In addition, each figure shows the data space position of the BMU corresponding to the sub-sample of data used for reconstruction. We see that the correlation of the stellar mass $\log(M_{*})$ and the absolute magnitude $M_{0.1 r}$ with the density field $\log(1+\delta)$ appear to be similar in different regions of the LSS. The upper two panels show reconstructions for sub-samples of heavy, red galaxies in high density regions classified as clusters (halos) according to the corresponding eigenvalues. The bottom-left panel belongs to heavy, red galaxies in low density regions classified as filaments and the bottom-right panel belongs to light, blue galaxies in regions classified as sheet (or filament since $\lambda_2 \approx 0$). Note that we adjusted the range of the y-axis in the last panel in order to improve the visibility of the correlation structure. Colors are defined according to the color classification code of \citet{2006MNRAS.368...21L}.}\label{fig:sdssup}
\end{figure*}

In order to compare the SOM to volume limitation, we train the SOM with the extended SDSS dataset (the SDSS properties including LSS properties) and select neuron-samples which appear to hold data close to the volume-limited samples in the $M_{0.1 r}$-$z$-plane. The SOM is set up to consist of $49$ neurons ($7 \times 7$ square lattice pattern) and the neuron-samples are generated as described in Section \ref{sec:som}. Note that for the training process all available quantities from the SDSS and the LSS are included. Therefore sampling is based not only on the flux-limitation bias encoded in the data distribution in the $M_{0.1 r}$-$z$-plane, but also takes into account additional systematics hidden in extra dimensions of the data sample. Figure \ref{fig:vol} shows the positions of all (left) and the selected (middle) trained neurons projected to the $M_{0.1 r}$-$z$-plane. In addition, we depict in the middle panels also the data samples corresponding to these selected neurons.

Furthermore the reconstructed correlation functions for the selected neuron-samples and the corresponding volume limited samples are shown for comparison on the right of Figure \ref{fig:vol}. For the lower magnitude (-18.5) sample the reconstruction appears to have a similar correlation strength compared to the neuron-sample. However, due to the fact that the neuron-sample includes galaxies above the magnitude limit introduced for volume limitation and not all galaxies below this threshold, the correlation functions appear to have an offset of $\approx 0.3$ orders of magnitude. For the higher absolute magnitude (-20.1) sample the reconstructed correlations differ more dramatically. As we will see in the next Section, the different correlation functions between magnitudes and the cosmic density are caused by additional systematics hidden in extra dimensions of the data. Those systematics are removed from sub-samples constructed by the SOM.

\subsection{SDSS application}\label{sec:sdssap}
As described in the previous Section, the application of the SOM to SDSS data results in various sub-samples of data holding different properties of the data space. Sub-samples can be used in order to reconstruct correlation functions between galaxy properties and the LSS. In addition, the data space weight of the corresponding neurons indicate the average properties of the galaxy sample. The reconstructed correlation functions for each sample as well as its average properties illuminate the relation of galaxies and their surrounding LSS. In order to illustrate this connection, we present reconstructions for various sub-samples in the following.

In particular, for correlation determination we include the r-band absolute magnitude $M_{0.1 r}$, the logarithm of the stellar mass $\log(M_{*})$ (in units of $M_{\mathrm{sun}} \ h^{-2}$) and the g-r color in the analysis and compare them to the logarithm of the cosmic large-scale density an a logarithmic scale $\log(1+\delta)$. Fig. \ref{fig:sdssup} shows the reconstructed correlation functions between galaxy properties and the density field.

\begin{figure*}[ht]
	\includegraphics[scale=1.2, angle=0]{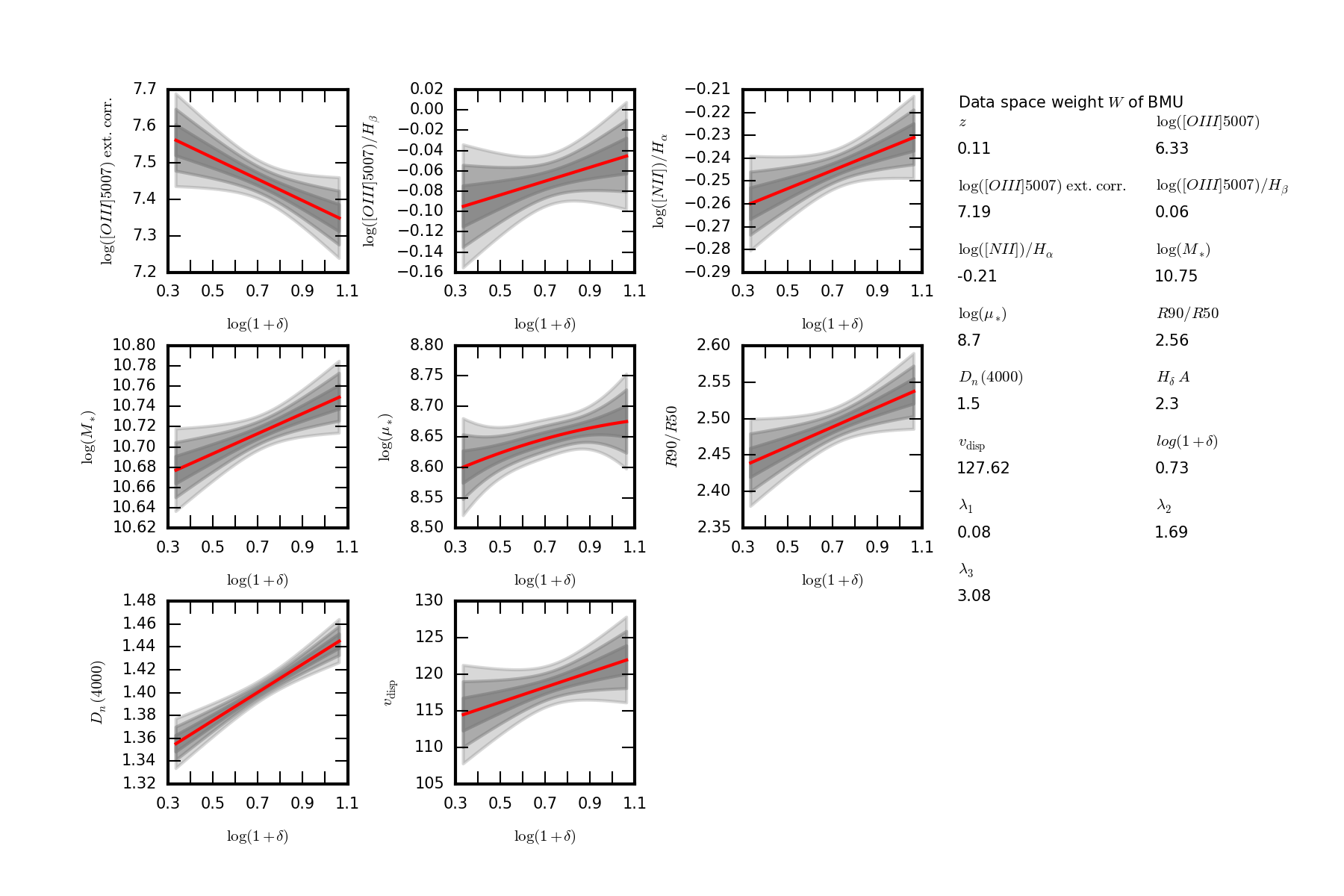}
	\centering
	\caption{Reconstructed correlation functions for one neuron-sample selected from the AGN data sample by the SOM. The data space position of the BMU is depicted in the right side of the figure.}\label{fig:agnup1}
\end{figure*}

\begin{figure*}[ht]
	\includegraphics[scale=1.2, angle=0]{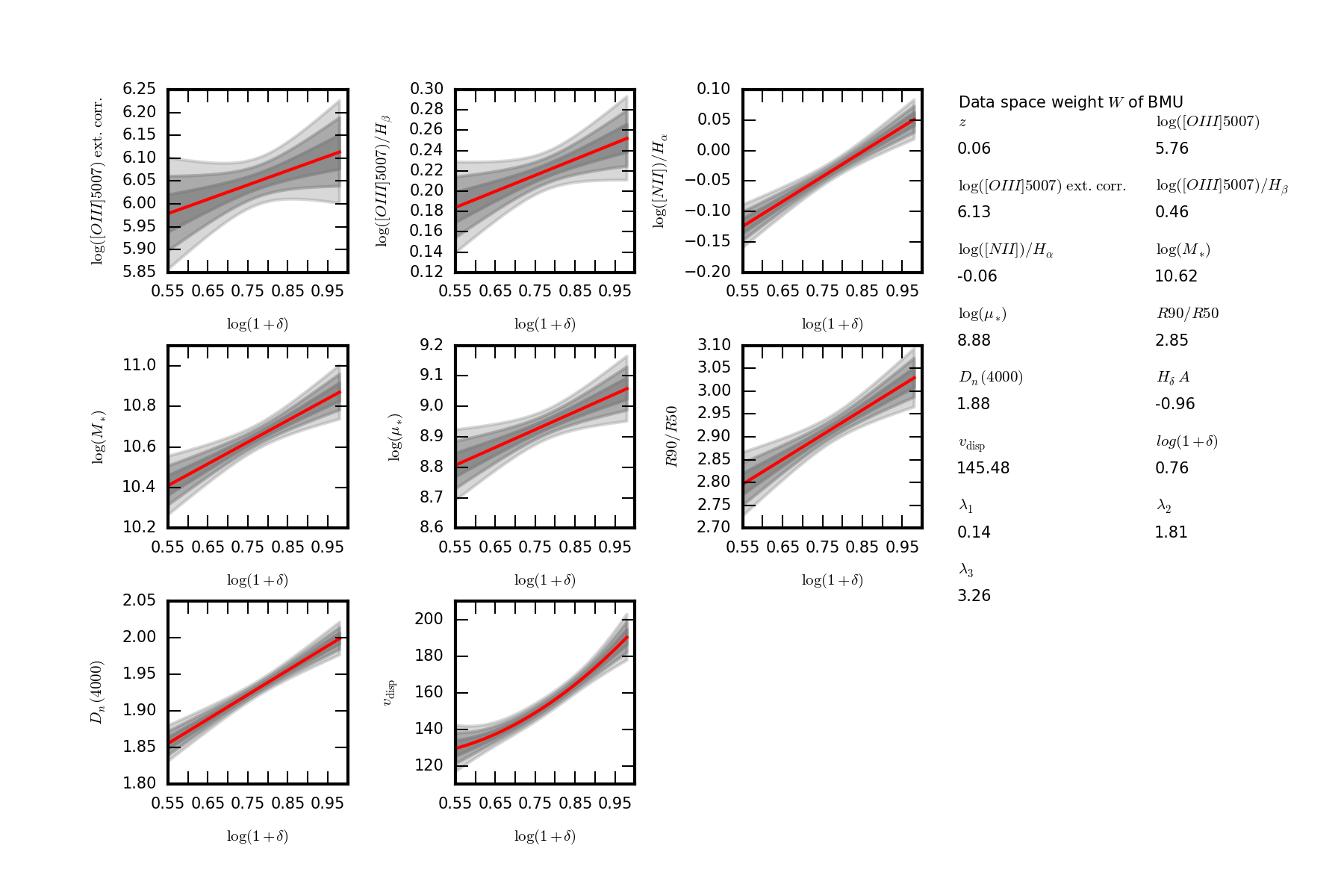}
	
	\centering
	\caption{Reconstructed correlation functions for one neuron-sample selected by the SOM from AGN data. }\label{fig:agnup2}
\end{figure*}

\begin{figure*}[ht]
	\includegraphics[scale=1.2, angle=0]{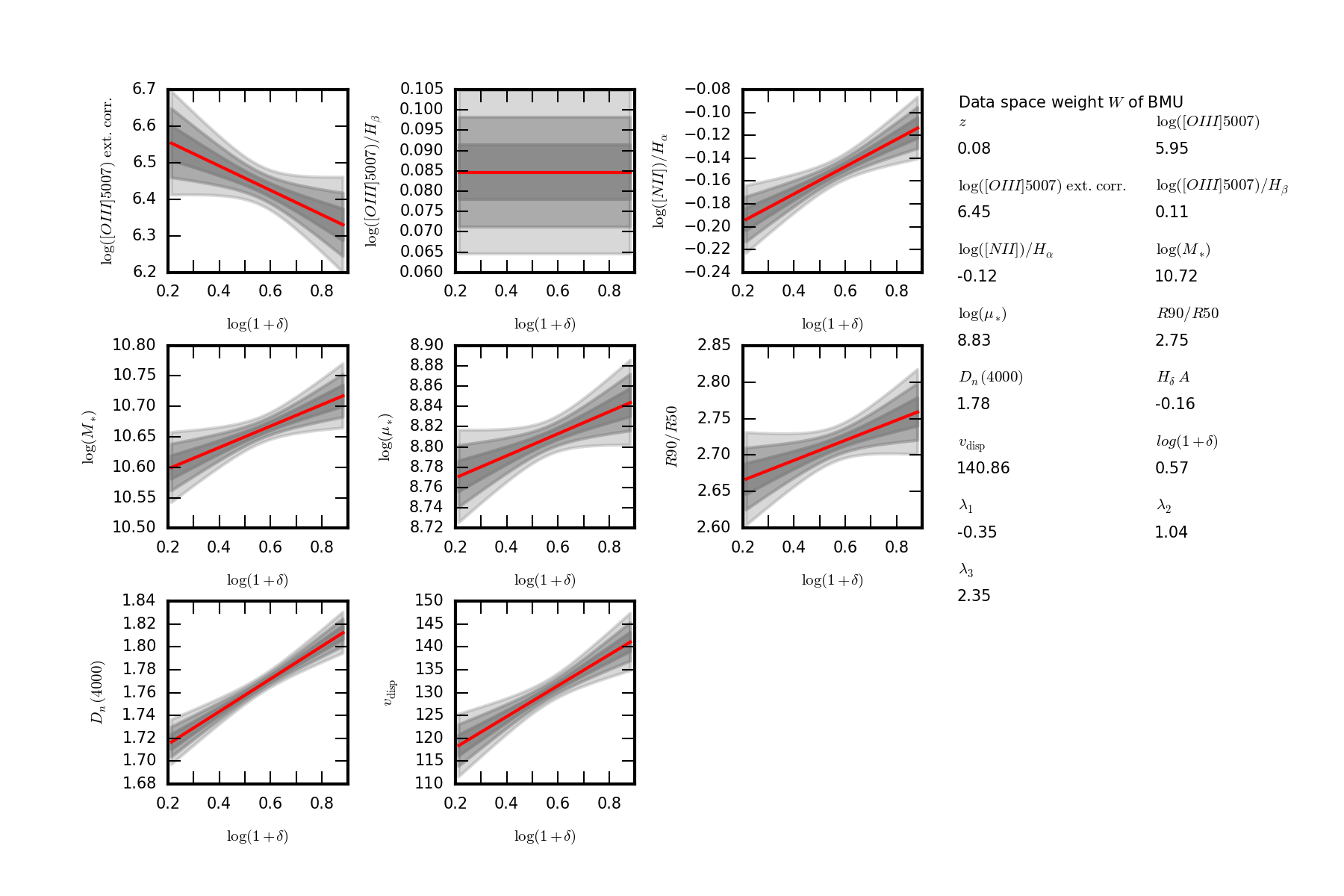}
	
	\centering
	\caption{Reconstructed correlation functions for one neuron-sample selected by the SOM from AGN data.}\label{fig:agnup3}
\end{figure*}

\begin{figure*}[ht]
	\includegraphics[scale=1.2, angle=0]{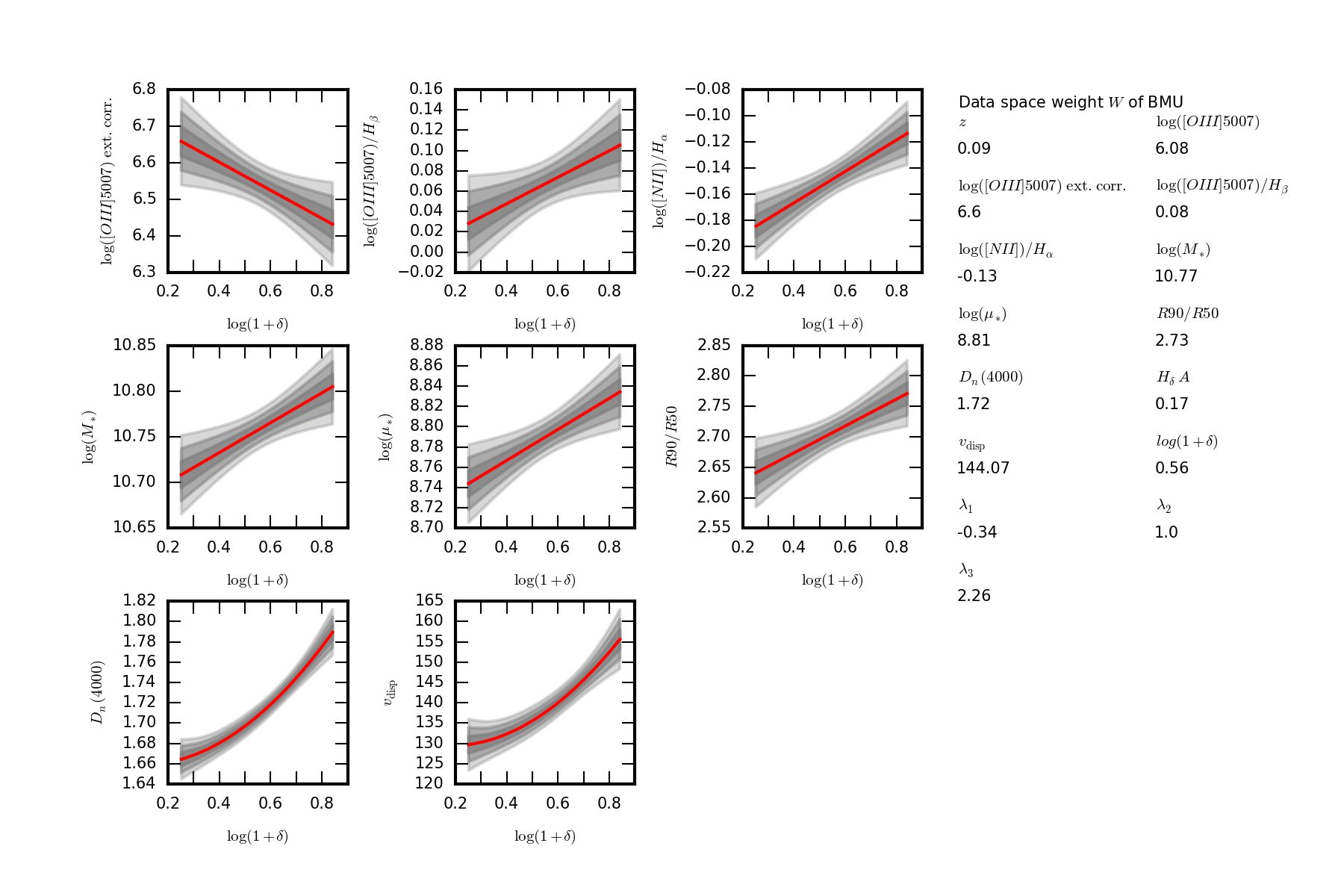}
	
	\centering
	\caption{Reconstructed correlation functions for one neuron-sample selected by the SOM from AGN data.}\label{fig:agnup4}
\end{figure*}

We see that the logarithm of the stellar mass appears to show a linear, positive correlation with the logarithm of the density field for multiple sub-samples. In particular, the upper two panels of Figure \ref{fig:sdssup} show reconstructions for sub samples of galaxies with the described density-mass relation. Both samples hold massive galaxies in a high density cosmic environment classified as cluster (halo) according to the eigenvalues of the tidal shear tensor. According to the color classification code described by \citet{2006MNRAS.368...21L}, both samples hold galaxies classified as red. Therefore we denote the samples as red samples in the following.

In addition, the SOM revealed a sub-sample of data (denoted as blue sample) holding blue galaxies of low mass in a low density cosmic environment classified as sheet (or filament since $\lambda_2 \approx 0.0$) depicted in the bottom-left panel of Figure \ref{fig:sdssup}. The masses of those galaxies appear to show a similar correlation with the density field compared to masses in red samples.

The visible positive correlation of stellar masses with their environmental matter density verify the intuitive conclusion that heavy galaxies are located in denser regions compared to light galaxies. However, it is of particular interest to point out that this trend is valid for light galaxies in low density regions (blue sample) as well as for heavy galaxies in high density regions (red samples).

In addition, the reconstructed correlation functions for the absolute magnitude show a linear dependency on the logarithm of the density. The results appear to be consistent with the correlations for stellar masses, since the brightness of galaxies in terms of the absolute magnitude is expected to be proportional to the logarithm of the stellar mass.

Correlations for colors indicate density dependence for blue and red galaxy samples.
In particular, we see that higher color values correspond to higher density regions on average, irrespective of the color classification of the sub-samples.

Our reconstructed correlations are consistent with the trends obtained by \citet{LEELI2008} in their studies of the correlations between physical properties of galaxies and the large scale environment. However, our recovered correlation amplitudes appear to differ from their results due to the fact that reconstructed amplitudes in the cosmic density field used by \citet{LEELI2008} are lower. The difference is caused by the fact that the reconstructions used by \citet{LEELI2008} have a larger voxel size ($\sim 6 \ \mathrm{Mpc \ h^{-1}}$) compared to the results of \citet{JASCHEBORG2012} ($\sim 3 \ \mathrm{Mpc \ h^{-1}}$). In addition, the BORG algorithm includes a more detailed treatment of uncertainties in the reconstruction.

In addition, SOMBI reveals existing systematics and unexpected correlation trends in the data. In particular, in the bottom-right panel of Figure \ref{fig:sdssup} we see inverted correlations in a sub-sample of data, compared to the correlations of the other (red and blue) samples. In total we identified 3 out of 49 sub-samples ($\approx 3\%$ of all data points) with similar data space weights as well as similar correlation structures. 
The representative sample holds heavy, red galaxies in low density regions located in filaments (or sheets, since $\lambda_2 \approx 0.09$). The reconstructed correlation seems to indicate that for this sample heavy galaxies appear to be located in lower density regions compared to light galaxies. In addition the color dependency on the density field disappears. A possible interpretation of the inverted correlation could be that in low density regions such as voids structures have formed a long time ago and therefore galaxies located in such regions are more likely to be old, red and heavy galaxies. In contrast, in high-density regions the increased presence of matter indicates an increased activity in galaxy and star formation. Therefore more young and light galaxies appear to be located in such regions. 

At this stage, our results are not capable of validating the described interpretation. The limiting factors are systematics caused by redshift distortions in the data sample. These distortions arise from peculiar velocities $\delta v$ of galaxies, which introduce a Doppler shift to the redshift measurement \citep[see e.g.][]{1987MNRAS.227....1K}. This effect causes galaxy clusters to appear stretched along the line of sight, an effect frequently referred to as ``Fingers of God''. The velocities of galaxies in high density regions rise up to $\delta v \sim 1000 \ \mathrm{km \ s^{-1}}$. Introducing a redshift uncertainty $\delta z \approx \delta v \ c^{-1}$ leads to uncertainties in the co-moving frame up to $\delta d_{\mathrm{com}} \approx 14 \ \mathrm{Mpc}$. Since the resolution of the BORG reconstruction maps is $\sim 3 \ \mathrm{M pc}$, a galaxy can be mapped 4 voxels away from its actual position in an extreme case. In addition, the BORG reconstructions are only corrected for redshift distortions up to linear order, but effects of non-linear redshift distortions may still be present in high density regions.

Therefore, at this point of the analysis we cannot verify whether the discovered sub-sample with inverted correlations indeed consists of heavy galaxies in low-density regions. Alternatively these could be galaxies actually located in high density regions but which are delocated by redshift distortions.

A more physical interpretation of galaxy physics is beyond the scope of this work and will be left for future publications.

\subsection{AGN application}\label{sec:agnap}
Now we apply SOMBI to galaxies classified as AGNs according to \cite{2003MNRAS.346.1055K}. The application follows the same strategy as described above resulting in $49$ sub-samples of data.

Galaxies hosting AGNs appear to have a higher stellar mass on average and are more likely to be located in higher density regions such as clusters (halos) or filaments. As a consequence of this all recovered sub-samples of data are located in filaments or in clusters.

As a preliminary consistency check we see that the reconstructed dependency of the stellar mass on the LSS density for all recovered sub-samples appears to be similar to the correlation structure of the full SDSS sample described above. Since the AGN data is a subset of data drawn from the SDSS sample, the correlation functions should be comparable.

In addition, in Figures \ref{fig:agnup1} - \ref{fig:agnup4} we present correlations for all available galaxy parameters of AGNs with the logarithm density field.
In particular, we reconstructed correlations for parameters associated with the recent star formation history and the LSS density field. We see that the $R90 / R50$ concentration index as well as the logarithm of the stellar surface mass density $\log(\mu_*)$ appear to be positive, linearly correlated to the logarithm of the density field. This result indicate that the star formation activity increases with increasing density, on average.

Structural parameters such as the $4000 \ \AA$ break strength $D_n (4000)$ as well as the intrinsic velocity dispersion $v_{\mathrm{disp}}$ of AGN galaxies appear to show a positive correlation with the logarithm of the density field. However, the revealed model for correlation (in particular the order of the polynomial as described in Section \ref{sec:BIC}) differs for various sub-samples. In particular, the resulting correlations for the velocity dispersion with the density appears to be linear for two sub-samples (Figure \ref{fig:agnup1} and \ref{fig:agnup3}) and follows a curved shape for the remaining sub-samples (Figure \ref{fig:agnup2} and \ref{fig:agnup4}).

The recovered correlation functions for structural parameters ($D_n (4000)$, $v_{\mathrm{disp}}$) as well as parameter associated with the recent star formation history ($R90 / R50$, $\log(\mu_*)$) show correlations with the density field consistent with the results obtained by \citet{LEELI2008}. As for the galaxy sample, correlation strengths differ compared to \citet{LEELI2008}.

Furthermore, we present correlations of the OIII 5007 and the NII emission line luminosities. The reconstructed correlation functions between the luminosity of the O III 5007 emission line ($\log([OIII] \ 5007)$ in solar units) and the density field appears to differ for the depicted sub-samples. In contrast, correlations for the OIII 5007 emission line relative to $H_{\beta}$ with the density field as well as correlations for the NII emission line relative to $H_{\alpha}$ with the density field appear to be more stable throughout depicted sub-samples. The results indicate that both, the OIII 5007 luminosity relative to $H_{\beta}$ and the NII luminosity relative to $H_{\alpha}$ increase with increasing density. A physical interpretation of these results is beyond the scope of this work.

We believe that the automatic classification of sub-samples of galaxies as well as the presented correlation analysis with the LSS is capable of revealing additional information about the connection between the LSS and galaxy properties. However, the goal of this work is to present the SOMBI method and to outline possible fields of application.

\section{Summary \& Conclusion}\label{sec:summ}
This work describes the implementation and application of the SOMBI algorithm, a Bayesian inference approach to search for correlations between different observed quantities in cosmological data. As an example we infer relations between various properties of galaxies and the cosmic large-scale-structure (LSS). This is of particular scientific interest, since the properties of galaxy formation and evolution are assumed to be directly linked to the LSS of our Universe. Studying the correlation between galaxies and their LSS environment will hence give further insight into the process governing galaxy formation.

Cosmological data generally consists of multiple sub-samples drawn from various different generation processes. Each sub-sample is expected to hold unique correlation structures. Therefore, for the SOMBI algorithm we seek to find a way to distinguish sub-samples of data belonging to different processes and to determine the correlation structure of each sample.

The correlation determination used by SOMBI assumes the correlation structures to be a polynomial with unknown order. The method infers a posterior PDF of the coefficients describing correlation via a Wiener Filter approach. To automatically choose the polynomial order, supported by the data, we employ a model selection method based on the ``Bayesian information criterion'' (BIC). The BIC compares the likelihood of different models matching the data. Apart from our initial restrictions and the restriction that data is drawn from a single generation process, this allows us to compare galaxy properties to properties of the LSS without prior information about correlation structures.

To ensure a successful application of the correlation determination method we automatically distinguish sub-samples of data belonging to different data generation processes. 
This is done by a specific kind of artificial neural network called ``Self Organizing Map'' (SOM). A SOM seeks to classify and distinguish sub-samples of data in noisy and highly structured observations. To do so, the SOM approximates the distribution of data by mapping a low-dimensional manifold (in this work two dimensional) onto the data space. The SOM provides sub-samples of similar data. We assume that each sub-sample consists of data drawn from one single generation process. To those samples the correlation analysis can then be applied successfully.

We test the performance of the SOMBI algorithm with mock data and cosmological data. For the latter we compare our results to simple volume-limitation sub-sampling.

As an illustrative example, we apply the SOMBI algorithm to two datasets, a galaxy and an AGN catalog based on the SDSS, in order to study the connection between galaxy and LSS properties. LSS information, as used in this work, is provided by the BORG algorithm \citep[see][]{JASCHEBORG2012}, a fully Bayesian inference framework to analyze 3D density fields in observations.

The application of the SOMBI algorithm to the described datasets shows that galaxy properties are clearly connected with the LSS. In particular, for the galaxy sample, stellar masses and absolute magnitudes appear to be linear, positive correlated to the cosmic density field on a logarithmic scale. In addition, we look at the revealed correlations of the color of galaxies and the LSS density. The reconstructed correlation functions imply that redder galaxies appear to be closer to dense regions.

Furthermore, we present correlations for additional galaxy properties such as structural parameters, parameters associated with the recent star formation history, velocity dispersions and luminosities of specific emission lines. Parameters are drawn from a subset of SDSS galaxies hosting AGNs. The results indicate that all described properties are correlated with the cosmic density field. However, correlation strengths appear to differ for recovered sub-samples, as classified by the SOM.

We conclude that the combined results ranging from the classification of galaxies according to data space properties to the revealed correlation structures revealed by the SOMBI algorithm provide insights into galaxy formation and evolution in specific cosmic environments on a preliminary level. A more detailed application of the SOMBI algorithm to cosmological data will be left for future work.

The generic framework of our method allows a simple analysis of many different kinds of datasets, including highly structured and noisy data. In addition, SOMBI is applicable for structure identification and correlation determination in different but related fields.

\begin{acknowledgements}
We thank Maksim Greiner and Fabian Schmidt for comments on the manuscript.
This research was supported by the DFG cluster of excellence "Origin and Structure of the Universe" (www.universe-cluster.de).

Funding for the Sloan Digital Sky Survey (SDSS) has been provided by the Alfred P. Sloan Foundation, the Participating Institutions, the National Aeronautics and Space Administration, the National Science Foundation, the U.S. Department of Energy, the Japanese Monbukagakusho, and the Max Planck Society. The SDSS Web site is http://www.sdss.org/. 
\end{acknowledgements}

\bibliographystyle{aa} % style aa.bst
\bibliography{SOMBI} % your references Yourfile.bib

\begin{appendix}
\section{PDF for realizations of reconstructed correlation functions}\label{sec:con}
For visualization the posterior of $\mathbf{f}$ (Eq. \eqref{eq:wienf}) can be transformed into data space resulting in a PDF for the realizations of the correlation function $f(x)$. In Eq. \eqref{eq:mod} we assumed that $f(x)$ can be Taylor expanded up to order $M$. Therefore the mean $\left<f(x)\right>$ is derived as:
\begin{align}\label{eq:ymean}
\bar{f}(x)&=\left<f(x)\right> \approx \mathbf{\tilde{R}}(x) \left<\mathbf{f}\right> = \mathbf{\tilde{R}}(x) \mathbf{f}_{\mathrm{WF}} = \sum\limits_{i=0}^{M} x^i \left<\mathbf{f}_i\right> \notag\\ &= \begin{pmatrix}
1,&x,&x^2,&... & x^M
\end{pmatrix}
\begin{pmatrix}
\left<\mathbf{f}_0\right>\\
\left<\mathbf{f}_1\right>\\
...\\
\left<\mathbf{f}_M\right>\\
\end{pmatrix}
\end{align}
with $x \in \mathbb{R}$ and $\mathbf{\tilde{R}}(x) : \mathbb{R}^{M+1} \rightarrow \mathbb{R}$. $\mathbf{\tilde{R}}$ has the same structure as $\mathbf{R}$ (Eq. \ref{eq:setup}) but the finite dimensional part of the operator, corresponding to the data points $x_i$, has been replaced by an infinite dimensional part for $x \in \mathbb{R}$.

Analogously we obtain the covariance $\mathbf{Y}$ as:
\begin{align}\label{eq:ycov}
\mathbf{Y}_{xy} &=\left<(f(x)-\bar{f}(x))(f(y)-\bar{f}(y))^T\right> \notag\\ &\approx \left<(\mathbf{\tilde{R}}(x)\mathbf{f}-\mathbf{\tilde{R}}(x)\mathbf{f}_{\mathrm{WF}})(\mathbf{\tilde{R}}(y)\mathbf{f}-\mathbf{\tilde{R}}(y)\mathbf{f}_{\mathrm{WF}})^T\right> \notag\\ &= \mathbf{\tilde{R}}(x)\left<(\mathbf{f}-\mathbf{f}_{\mathrm{WF}})(\mathbf{f}-\mathbf{f}_{\mathrm{WF}})^T\right>\mathbf{\tilde{R}}(y)^T \notag\\
&= \mathbf{\tilde{R}}(x)\mathbf{D}\mathbf{\tilde{R}}(y)^T \notag\\
&= p_* \mathbf{\tilde{R}}(x)(\mathbf{R}^T\mathbf{R})^{-1}\mathbf{\tilde{R}}(y)^T
\end{align}
Combining these results yields a PDF for the possible realizations of the fitted curve
\begin{equation}
P(f(x)|\mathbf{d})=\mathcal{G}(f(x)-\mathbf{\tilde{R}}(x)\mathbf{f}_{WF},\mathbf{Y}) \ ,
\end{equation}
which describes how likely a realization is, given the data. This permits to visualize the fitted curve including corresponding uncertainties in specific areas of the data space.

\section{Self Organizing Map algorithm}\label{sec:apsom}
Various different implementations of SOMs have been presented in the literature \citep[see e.g.][]{2001som..book.....K}. Many implementations appear to follow the same generic idea but differ in some implementation details. The difference is caused by the fact that SOMs are used in order to tackle many different questions regarding the structural form of data. Therefore, we present the detailed implementation of our SOM algorithm in the following.

A SOM is an artificial neural network specifically designed to determine the structure of datasets in high dimensional spaces. The network has a specific topological structure. In this work we rely on a network with neurons interlinked in a square-lattice pattern with a neighbourhood function representing the strength of those links. The network is trained by data with a training algorithm which gets repeated for every data point multiple times resulting in a learning process. The generic form of the network as well as the learning process is described in Section \ref{sec:som}.

Before the process can start the network has to be linked to data space. Therefore each neuron holds a vector $\mathbf{W} =(W_1 , W_2 , ... , W_N)^T$ in the $N$ dimensional data space, called weight. It is important to point out that the neurons live in two different spaces: the data space with the position represented by its weight and the network pattern where each neuron is linked to each other by a neighbourhood function.

In the beginning of the learning process, no information about the data space has been provided to the network. Therefore weights are initialized randomly in data space. After initialization the actual learning process starts.
Each iteration of the learning process follows the same generic form.

First the ``Best matching unit" (BMU) is calculated for a randomly chosen data vector $\mathbf{V}= (V_1 , V_2 , ... , V_N)^T$. The BMU is defined to be the closest neuron to $\mathbf{V}$ in terms of similarity, as expressed by a data-space distance measure. For this we use the Euclidean distance $D$ in rescaled data dimensions. Specifically
\begin{equation}
D=\sqrt{\sum\limits_{i=1}^N \left(\frac{V_i-W_i}{\sigma_i}\right)^2} \ ,
\end{equation}
where $\sigma_i$ being the scale factor for each component $i$. This automatically solves the problem to compare quantities with disparate units. We define $\sigma_i$ as:
\begin{equation}
\sigma_i:=V_{i \ \mathrm{max}} - V_{i \ \mathrm{min}} \ ,
\end{equation}
where $V_{i \ \mathrm{max}}$ and $V_{i \ \mathrm{min}}$ are the maximum and minimum values of the $i$th component of all data vectors.

The weight of the neuron for which $D$ gets minimal is modified according to the value of $\mathbf{V}$. Therefore the new weight for the BMU at iteration step $t+1$ is:
\begin{equation}
\label{eq:BMU}
\mathbf{W}_{t+1}=\mathbf{W}_t+L_t(\mathbf{V}-\mathbf{W}_t) \ ,
\end{equation}
where $\mathbf{W}_t$ is the previous weight and $L_t$ is the ``learning rate". The learning rate is a decreasing function of $t$ and hence quantifies how strong an input vector should influence the weights at a specific iteration step. It has to be a decreasing function since the $t$th vector presented to the network should not change the weight of a neuron as much as the previous ones to ensure converging information updates. There are two convenient shapes for learning rates: a linear and an exponential decay. In this work we chose to use the exponential decay with $L_t$ given as:
\begin{equation}
\label{eq:LR}
L_t=L_0 e^{-\frac{t}{\lambda}} \ .
\end{equation}
$L_0$ is the initial learning rate and $\lambda$ is a tunable parameter to adopt the change of the learning rate for each iteration.

Since neurons are linked to each other, adaptation of individual neurons will also affect the weights of all other neurons. The strength of the modification of those weights should decrease with distance to the BMU in the specified topology of the network. Therefore the size of the neighbourhood of a single neuron for a specific iteration step $t$ is
\begin{equation}
\sigma_t= \sigma_0 e^{-\frac{t}{\lambda}} \ ,
\end{equation}
where $\sigma_0$ is the initial neighbourhood size. Note that the size decreases with $t$ in order to ensure that the modification of the vicinity of the BMU gets less important with increasing $t$. The neighbourhood size $\sigma$ defines the influence rate $\Theta$ of one iteration:
\begin{equation}
\label{eq:inflr}
\Theta_t= e^{-\frac{d_{\mathrm{BMU}}^2}{2 \sigma_t^2}} \ ,
\end{equation}
where $d_{\mathrm{BMU}}$ is the distance between the position of the updated neuron and the BMU of the $t$th iteration step in the square lattice pattern. It is important to distinguish $d_{\mathrm{BMU}}$ from $D$, since $d_{\mathrm{BMU}}$ is the distance between two neurons in the network pattern and $D$ is the euclidean distance in data space. Note that $\Theta$ assumes a value of one for the BMU itself therefore modification functions can be combined, yielding
\begin{equation}
\mathbf{W}_{t+1}=\mathbf{W}_t+L_t \Theta_t (\mathbf{V}-\mathbf{W}_t) \ .
\end{equation}
These steps are repeated for every single vector in the dataset. 

To avoid biasing weights to the first subset of data, the whole learning process has to be repeated multiple times. The final result of the learning process is given by averaging the weights for each learning process.

\section{Mapping the SDSS data onto reconstructed density fields}\label{sec:mapping}

In order to extract the properties of the density field reconstructions from the results provided by the BORG algorithm, we map the SDSS data onto the reconstructed volume.
More precisely, we look for the position of each galaxy in the cubic volume and store the properties of the LSS in the voxel hosting the galaxy. All galaxies within one voxel are assigned the same LSS information. This results in an extended data catalog, containing the intrinsic properties of the galaxies as well as the properties of the LSS in the surrounding area of each galaxy. Note that this procedure is perfectly applicable for all kinds of cosmological data as long as there is information about the 3D position of the objects in the data.

Since the SDSS data provides position information in redshift space, we need to transform the coordinates to the co-moving frame. Redshifts $z_i$ are transformed to co-moving distances $d_{\mathrm{com}}$ according to:
\begin{equation}
d_{\mathrm{com}}= \int_{0}^{z_i} \frac{1}{c H(z)} \mathrm{d}z \ ,
\end{equation}
where $c$ is the speed of light and $H(z)$ denotes the Hubble parameter. $H(z)$ is given as:
\begin{equation}
H(z)= H_0 \sqrt{\Omega_m (1+z)^3+\Omega_c (1+z)^2+\Omega_{\Lambda}} \ ,
\end{equation}
under the assumption of a concordance $\Lambda$CDM model with the cosmological parameters $\Omega_m=0.24$, $\Omega_c=0.00$, $\Omega_{\Lambda}=0.76$, $h=0.73$ and $H_0=h \ 100 \ \mathrm{km \ s^{-1} \ MPc^{-1}}$ \citep[see][]{2007ApJS..170..377S}. We used this set of parameters instead of more recent ones in order to match the cosmology used for the LSS reconstructions.

As a final step we calculate the Cartesian coordinates for each galaxy:
\begin{equation}
x=d_{\mathrm{com}} \cos(\delta) \cos(\alpha)
\end{equation}
\begin{equation}
y=d_{\mathrm{com}} \cos(\delta) \sin(\alpha)
\end{equation}
\begin{equation}
z=d_{\mathrm{com}} \sin(\delta) \ ,
\end{equation}
where $\alpha$ and $\delta$ are the right ascension and declination of the ecliptic frame, respectively.

Since the BORG reconstruction maps provide an approximate PDF for the density field, we see that uncertainties in the reconstruction increase with increasing distance. Therefore, in order to exclude areas of high uncertainties in the analysis of correlation determination, we excluded all galaxies of the SDSS sample above a certain distance $d_{\mathrm{lim}}=450 \ \mathrm{Mpc \ h^{-1}}$. This results in a sub-sample including only galaxies with redshifts between $0.001<z<0.156$. Due to the fact that the BORG reconstruction maps are based on the SDSS, uncertainties in the reconstruction increase in regions with less signal, specifically regions with a low number of galaxies. Therefore the majority of data remains included in the limited sample.

\section{Probability distribution for correlation functions with the LSS}\label{sec:gm}
As described in Section \ref{sec:borg} the BORG algorithm provides an ensemble of density contrast field realizations that capture observational uncertainties. In order to treat the uncertainties in the density contrast correctly during correlation determination, the reconstruction algorithm described in Section \ref{sec:param} has to be applied to each realization independently. This yields a PDF $P(\mathbf{f}|\delta_i \mathbf{d})$ for each $\delta_i$. The dependency of the realizations has to be marginalized out of the PDF's in order to obtain the final PDF for the correlation function $P(\mathbf{f}|\mathbf{d})$. This results in a Gaussian mixture for the posterior PDF. Specifically,
\begin{equation}
\label{eq:marg}
\begin{split}
P(\mathbf{f}|\mathbf{d})=\int P(\mathbf{f},\delta|\mathbf{d}) \mathrm{d}\delta=\int P(\mathbf{f}|\delta,\mathbf{d}) P(\delta|\mathbf{d}_*) \mathrm{d}\delta
\\ \approx \frac{1}{S} \sum\limits_{i=1}^S \delta^D(\delta-\delta_i) P(\mathbf{f}|\delta_i,\mathbf{d}) = \frac{1}{S} \sum\limits_{i=1}^S \mathcal{G}(\mathbf{f}-\mathbf{m}_i,\mathbf{D}_i) \ ,
\end{split}
\end{equation}
where $\delta_i$ denotes one of the $S$ realizations of the density contrast and $\mathbf{m}_i$ and $\mathbf{D}_i$ denote the corresponding mean and covariance for each fit.

\end{appendix}

\end{document}